\documentclass{paper}
\usepackage{graphicx}
\title{Non-Abelian Einstein-Born-Infeld-Dilaton Cosmology}
\author{A. F\"uzfa$^{1}$, J.-M. Alimi$^{2}$, \\
               Laboratory Universe and Theories, CNRS UMR 8102,\\
 Observatoire
       de Paris-Meudon and Universit\'e Paris 7, France\\
           $^{1}$ email:  andre.fuzfa@obspm.fr\\
           $^{2}$ email: jean-michel.alimi@obspm.fr}
\begin{document}
\maketitle
\label{firstpage}
\begin{abstract}
The non-abelian Einstein-Born-Infeld-Dilaton theory, which rules the dynamics
of tensor-scalar gravitation coupled to a $su(2)$-valued gauge field ruled by Born-Infeld lagrangian,
is studied in a cosmological framework.
The microscopic energy exchange between the gauge field and the dilaton
which results from a non-universality of the coupling to gravity modifies the usual behaviour of tensor-scalar theories coupled to matter fluids.
General cosmological evolutions are derived
for different couplings to gravitation and a comparison to universal coupling is highlighted.
Evidences of cosmic acceleration are presented when the evolution is interpreted in the Jordan physical frame
of a matter respecting the weak equivalence principle. The importance for the mechanism of cosmic acceleration
of the dynamics of the Born-Infeld gauge field, the attraction role
of the matter fluid and the non-universality of the gravitational couplings is briefly outlined.
\end{abstract}
\section{Introduction}
In the description of the very beginning of the universe, well before the Big
Bang nucleosynthesis, field theoretical
models are to be considered instead of the usual hydrodynamical
description of matter. Those kinds of models, inspired by high-energy
physics, have lead to numerous progress in modern cosmology, trying to
solve various problems from cosmic acceleration or the flatness
problem to the magnitude of the cosmological constant or the existence
of topological defects. The motivation of this paper lies in two
questions, amongst many others, that raise from the description of the
very first moments by high energy physics.\\
\\
First is the question whether large scale massless gauge fields
can play any interesting role in cosmology. Indeed, such fields
could have existed in the early universe before the phase transitions of
spontaneous symmetry breaking but, if they were ruled by usual
Yang-Mills (YM)
conformally invariant dynamics, their primeval excitations have
probably been
swept away by inflation. This point motivated some authors to study
the impact of the Born-Infeld (BI)
type modification of gauge dynamics, suggested by string theory, on
cosmology (see \cite{dyadichev}). The BI lagrangian breaks down the
scale invariance of the gauge fields beyond some critical energy, and
therefore it is not obvious to conclude directly on the becoming of such gauge
fields during and after an inflation period. Furthermore, it was
proved in \cite{dyadichev} that such gauge fields of BI-type cannot
provide any cosmic
acceleration on their own although they can mimic a fluid of negative
pressure. Before going further on this question, it is therefore of
first importance to study more deeply in a cosmological context the interaction between gauge
and scalar fields as suggested in models inspired by high-energy physics.
\\
The second question is to see what happens to a possible scalar sector of
gravitation during the cosmological evolution.
Indeed, string theories predict the existence of the dilaton \cite{witten}, a Lorentz scalar
partner to the tensor Einstein graviton as low-energy limit of
bosonic actions. This large theoretical
framework provides a physical background for tensor-scalar
modification \cite{jordan} of general relativity, in which gravitation is mediated
by long-range scalar field acting in complement of the usual spin 2
gravity fields.
Although this question has been widely studied
when dilaton - or more generally
tensor-scalar theories - is coupled to matter during radiation and
matter dominated era, the case of a microscopic field model which would not be coupled universally to
gravity, as suggested in string theory, has been less considered.
In particular, how does the interaction between scalar and gauge fields
modify their respective dynamics and the resulting cosmological
evolution will be the main subject of the present paper.\\
\\
But before going any further, let us locate the present work in the
existing litterature. In this paper, we will focus on cosmological
solutions of the Einstein-Born-Infeld-Dilaton (EBID) equations for flat
spacetimes.
Cosmologies with large scale massless homogeneous and isotropic gauge
fields with gauge group $SU(2)$ and ruled by usual YM dynamics have
been studied for a long time
(\cite{henneaux,galtsov,cervero,hosotani,demaret}).
The gravitational instability of flat spacetimes filled with such
gauge fields was studied in \cite{fuzfa}. Generalisation to higher gauge
groups have also been studied \cite{bertolami,moniz} in the case of flat and
closed cosmologies.
The Einstein-Born-Infeld cosmology with non-abelian gauge fields
deriving from gauge group $SU(2)$ has been studied thoroughly in
\cite{dyadichev} for flat, closed and open spacetimes for any value
of the cosmological constant. The minimal coupling of
large scale cosmological gauge fields and scalar multiplets has been studied
in \cite{moniz2,henriques}. The Einstein-Yang-Mills-Dilaton (EYMD) equations
for flat cosmologies and a special case of non-universal coupling to gravity,
have been derived in \cite{bento,bento2} where the authors
highlighted the energy exchange between the
dilaton and the gauge fields and briefly discussed its effects on inflation,
entropy crisis and the Polonyi
problem (domination of a nearly massless dilaton at late
times). However, they did not propose a complete solution to the EYMD
field equations in the cosmological framework which could allow to
address completely these issues. Moreover, they did not discuss the general influence of non-universal coupling to gravity as well.
The EYMD system was also studied in
\cite{scialom} in the case of closed
Friedmann-Lemaitre-Robertson-Walker (FLRW) with a static gauge
field\footnote{Due to this particular topology, the EYMD system does not
  reduce to pure Einstein-Dilaton field equations when the gauge field is static.}
and vanishing dilaton potential and cosmological constant.\\
\\
As we told before, tensor-scalar cosmologies have been widely studied,
with a large spectrum of applications for physical cosmology:
inflation, primordial nucleosynthesis, Cosmic Microwave Background, ...
The question of the convergence of tensor-scalar theories to general
relativity during the cosmological evolution has been widely studied in
\cite{damour,alimi,barrow,serna} and references therein.
For the so-called ``\textit{Einstein conformal frame}'', where the
gravitational and scalar fields have pure spin 2 and spin 0 dynamics respectively,
the scalar sector of gravitation disappears naturally during cosmic expansion due to its coupling to matter.
The cosmological evolution of the dilaton emerging from string theory
has been studied in \cite{damour2}.\\
\\
In this paper, we will consider the cosmological evolution of the
dilaton coupled directly to a large-scale non-abelian gauge
field ruled by BI dynamics which go beyond the scale invariance of YM theory.
A non-universal coupling to gravity, as suggested in preceding works, will lead to quite different results
to the usual
coupling of tensor-scalar theories to a fluid.
For example,
when the gauge fields are
governed by YM scale-invariant dynamics, the scalar sector of gravitation remains
directly coupled to the gauge fields although they mimic a radiation
fluid. However, it is well-known that tensor-scalar theories decouple from
radiation (except during phase transition). Through both numerical
computations and analytical solutions, we will show how the dilaton
evolution is modified by non-universal coupling to the metric. This will lead to remarkable consequences
for cosmology.
\\
\\
The structure of this paper will be as follows. In section 2, we
establish the general field equations for the EBID cosmology.
In section 3, we first remind the reader about BI cosmology as was studied in
\cite{dyadichev}.
The non-abelian BI cosmology can be split in two extreme regimes depending on the energy density of
the gauge field compared with a critical scale introduced in the BI
theory. For large energies, the gauge field is shown to mimic a fluid
of negative pressure with $p/\rho=-1/3$ while in the low-energy limit
the scale-invariant YM dynamics is retrieved and the gauge field looks
like a radiation fluid.
On the other hand, we also remind the reader about dilaton cosmology, studied in
\cite{damour,damour2}. As the equations
are to be established and solved with pure spin degrees of freedom in
the Einstein frame, the interpretation in terms of the Jordan physical
frame is also recalled. The case of universal coupling is solved there with
usual properties of tensor-scalar theories.
In section 4, we will focus on the strong field limit of the EBID
system with general coupling to gravity and in section 5 the low-energy limit which consists of a generalised version
of the EYMD system appearing in \cite{bento,bento2} is treated.
In section 6, we analyse a complete cosmological evolution of the EBID
system and present evidences for possible cosmic acceleration in the
physical Jordan frame. This frame is defined with respect to a pressureless matter fluid that has
been added to the gauge sector. The acceleration is shown to resist to the attraction provided
by matter and appears to be intrinsically related to the non-universality of the coupling to gravity.
Finally, we conclude in section 7 by some
perspectives to the present work.
\section{Field Equations of Einstein-Born-Infeld-Dilaton
  Cosmology}
Most of the interest of field models, for example
deriving from string theory in
the low energy-limit, comes from their non-universal coupling to the
gravity fields $g_{\mu\nu}$ and its scalar counterpart $\phi$
(each  type of matter field has in general its own coupling
function to the dilaton, see \cite{damour2}). This results in a violation
of the weak equivalence principle and therefore the gravitational interaction
of these microscopic field models is different
from an usual tensor-scalar theory where the weak equivalence principle is
usually assumed.
Without imposing such a violation of the weak equivalence principle
at a microscopic scale, field models would not be different than
considering a tensor-scalar theory in presence of a fluid
with the equation of state of the considered fields.
Therefore, we will make use of a general form of the action for the
non-abelian Einstein-Born-Infeld-Dilaton system,
that takes into account a possible violation of the weak equivalence principle.
This action writes down
\begin{eqnarray}
\label{lebid}
S&=&\int\left\{-\frac{1}{2\kappa}R-\frac{1}{2}\partial^\mu \phi\partial_\mu
  \phi-V(\phi)-A^4\left(\phi\right)\mathcal{L}_{BI}\left(B^2(\phi)g_{\mu\nu},A_\mu\right)\right\}\sqrt{-g}d^4x\nonumber\\
&&+S_m\left[\psi_m,C^2(\phi)g_{\mu\nu}\right]\cdot
\end{eqnarray}
In this action, the gravitational interaction is described by the scalar curvature
$R$ and the dilaton $\phi$, $\kappa$ being the ``\textit{bare}'' gravitational coupling
constant and $A(\phi)$, $B(\phi)$ and $C(\phi)$ being three different coupling functions of the dilaton to matter.
The first two illustrate the coupling of the gauge sector
to the volume form and to the Einstein metric $g_{\mu\nu}$ inside the Born-Infeld lagrangian $\mathcal{L}_{BI}$
(where $A_\mu$ are the non-abelian gauge potentials) and the last coupling function $C(\phi)$ is related to another type of matter
ruled by the action $S_m$.
Another parametrisations
of Einstein-Born-Infeld-Dilaton action were considered in the
litterature, for example with $A=1$ and $B=\exp(k/2\phi)$
in \cite{bento,bento2,clement,scialom}
and with $B=A^2=\exp(k/2\phi)$ in \cite{tamaki}.
The non-abelian gauge interaction is
described by the Born-Infeld lagrangian build upon the field strength
tensor
$$
F_{\mu\nu}=F_{\mu\nu}^{\bf{a}}T_{\bf{a}},
$$
(with $T_{\bf{a}}$ are the generators of the gauge group under
consideration\footnote{Gauge indices will be noted as bold latin
letters.}) and its dual tensor $\tilde{F}_{\mu\nu}$. Indeed, this
lagrangian, denoted by $\mathcal{L}_{BI}$ in (\ref{lebid}), is
defined as
\begin{eqnarray}
\mathcal{L}_{BI}&=&\epsilon_{c}\left(\mathcal{R}-1\right)\nonumber\\
&=&\epsilon_{c}\left(\sqrt{1+\frac{B^{-4}(\phi)}{2\epsilon_{c}}F_{\mu\nu}F^{\mu\nu}-\frac{B^{-8}(\phi)}{16\epsilon_c^2}\left(F_{\mu\nu}\tilde{F}^{\mu\nu}\right)^2}-1\right)\cdot
\end{eqnarray}
where $\epsilon_c$ is the Born-Infeld critical energy and $B(\phi)$
is the dilaton coupling function which is equal to $e^{k/2\phi}$
when non-perturbative effects are not taken into account (see
\cite{bento,bento2,damour2}). In this case, $k$ will be called the
dilaton coupling constant. Throughout this paper, we will assume the
Planck system of units, in which $\hbar=c=1$ and $G=m_{Pl}^{-2}$,
with the Planck mass $m_{Pl}=1.2211\times 10^{19}GeV$ and the
gravitational coupling constant is $\kappa=8\pi G$. We have also set
the gauge coupling constant to unity, as it actually defines a
system of units for the dilaton field $\phi$, provided the dilaton
being massless ($V=0$). The Born-Infeld critical energy
$\epsilon_{c}$ defines the scale above which non local effects of
string theory arise and where the scale invariance of the gauge
fields is broken. For example, in the low-energy limit
$\epsilon_c\rightarrow\infty$ of the Born-Infeld part of the action
(\ref{lebid}), we recover the usual, conformally invariant,
Yang-Mills (YM) lagrangian density for the non-abelian gauge field:
\begin{equation}
\label{lym}
\sqrt{-g}\mathcal{L}_{YM}=-\sqrt{-g}A^{4}(\phi)B^{-4}(\phi)\frac{1}{4}F_{\mu\nu}^{\bf{a}}F^{\mu\nu}_{\bf{a}}\cdot
\end{equation}
In this paper, we will focus on cosmology and therefore we will adopt
the prescriptions of the cosmological principle which states that the
spatial sections of our Universe are homogeneous and isotropic.
For the sake of simplicity, we will also restrict ourselves to the
case of flat space-times, which constitutes however a very nice
approximation of the present universe and its early stages as well.
The metric describing such spacetimes is the one of
Friedmann-Lemaitre-Robertson-Walker (FLRW):
\begin{equation}
ds^2=-N^2(t) dt^2+a^2(t)\left(dr^2+r^2d\theta^2+r^2\sin^2\theta d\varphi^2\right)
\end{equation}
where $a(t)$ is the scale factor and $N$ is the so-called \textit{lapse function} of the hamiltonian
Arnowitt-Deser-Misner approach to general relativity. This function
can be fixed by a specific choice of time coordinate (gravitational
gauge freedom). The symmetries prescribed by the cosmological
principle we assume impose that the dilaton scalar field $\phi$
depends only on time.\\
\\
A remarkable fact about non-abelian gauge fields is that they admit
non-trivial homogeneous and isotropic configurations at the opposite
of their abelian $U(1)$ counterparts (see \cite{dyadichev,galtsov} and
references therein for
a complete discussion of the $SU(2)$ case). The main reason for that is because only the
gauge invariant quantities such as the field strength tensor have to
exhibit the symmetries explicitly, while the gauge potentials can be
symmetric up to a gauge transformation (see \cite{forgacs, henneaux}
for more general gauge groups).
As a result, the energy can be distributed amongst the different gauge
degrees of freedom while the stress-energy tensor remain compatible
with the maximal symmetry of the space-time background. In this paper,
we will restrict ourselves to the case of $su(2)-$valued gauge
potentials, for which the ansatz
\begin{equation}
\label{FLRWamu}
{\bf{A}} =A_\mu^{\bf{a}} T_{\bf{a}}dx^\mu=\sigma(t) T_{\bf{m}}dx^{\bf{m}}
\end{equation}
of the connexion one-form $\bf{A}$ makes the gauge invariant
quantities satisfying the required symmetry (see \cite{henneaux,galtsov}). The remaining dynamical degrees of freedom
of the gauge potential are now expressed by the field $\sigma(t)$.
However, our results will
not depend on this particular choice as the ansatz above can be
generalised to higher gauge groups (see \cite{bertolami,moniz}).
In the equation above,
the generators $T_{\bf{m}}$ of the Lie algebra of the gauge group
$SU(2)$ are to be taken in the coordinate dependent basis of the gauge
degrees of freedom space as follows:
\begin{eqnarray}
T_{\bf{r}}&=&\sin\theta\cos\varphi T_{\bf{1}}+\sin\theta\sin\varphi T_{\bf{2}}
+\cos\theta T_{\bf{3}}\nonumber\\
T_{\bf{\theta}}&=&\frac{\partial}{\partial \theta}T_{\bf{r}}\nonumber\\
T_{\bf{\varphi}}&=&\frac{\partial}{\sin\theta\partial \varphi}T_{\bf{r}}\nonumber
 \end{eqnarray}
with $T_{\bf{i}}=\frac{1}{2}\sigma_i$ the usual basis of the Lie
algebra $su(2)$ ($\sigma_i$ being the Pauli matrices) with the
following standard normalization conditions and commutation relations:
$$
tr\left( T_{\bf{a}}T_{\bf{b}} \right)=\frac{1}{2}\delta_{\bf{a}\bf{b}}, \; \left[T_{\bf{a}},T_{\bf{b}}\right]=i\epsilon_{\bf{ab}}^{\bf{c}}T_{\bf{c}}
$$
The ansatz (\ref{FLRWamu}) is of course independent of the particular choice of the
action for the gauge field, as was shown in \cite{dyadichev}.\\
\\
The symmetries implied by the cosmological principle therefore allow
us to write down
(\ref{lebid}) as an effective one dimensional action, after integrating over
$R^3$ and dividing by the infinite volume of its orbits:
\begin{equation}
\label{ebid}
S_{eff}=\int dt\left\{ -\frac{3}{\kappa}\frac{\dot{a}^2a}{N}+\frac{\dot{\phi}^2}{2}\frac{a^3}{N}-V(\phi)Na^3-Na^3\epsilon_{c}A^4(\phi)\left(\mathcal{R}-1\right)\right\}+S_m
\end{equation}
where a dot denotes a derivative with respect to the time $t$ and
where $\mathcal{R}$ is given by
\begin{equation}
\label{bigr}
\mathcal{R}=\sqrt{1-3\frac{B^{-4}(\phi)}{\epsilon_{c}}\left(\frac{\dot{\sigma}^2}{a^2N^2}-\frac{\sigma^4}{a^4}\right)-9\frac{B^{-8}(\phi)}{\epsilon_c^2
  a^6 N^2}\dot{\sigma}^2\sigma^4},
\end{equation}
where $\sigma(t)$ is the gauge potential as defined in (\ref{FLRWamu}).
Following \cite{dyadichev}, it is also convenient to write $\mathcal{R}$ as
\begin{equation}
\mathcal{R}=\sqrt{1-\Gamma}\sqrt{1+\Delta}
\end{equation}
with
\begin{equation}
\label{gamdel}
\Gamma=\frac{3\dot{\sigma}^2 B^{-4}(\phi)}{\epsilon_{c} a^2 N^2}\; ,\; \Delta=\frac{3\sigma^4 B^{-4}(\phi)}{\epsilon_{c} a^4}\cdot
\end{equation}
From the action (\ref{ebid}) and relation (\ref{bigr}), it is straightforward to write down the field equations for the
Einstein-Born-Infeld-Dilaton system by varying this action over the
following degrees of freedom: $N,a,\phi$ and $\sigma$. First, the Euler-Lagrange
equation for the variable $N$ gives the hamiltonian constraint
\begin{equation}
\label{hebid}
\left(\frac{\dot{a}}{a}\right)^2=\frac{\kappa}{3}\left[\frac{\dot{\phi}^2}{2}+V\left(\phi\right)+\epsilon_{c}A^4(\phi)\left(\mathcal{P}-1\right)+\rho_m\right]
\end{equation}
which we will refer to as the $Friedmann$ equation. In the previous
equation, $\rho_*$ stands for the energy density of the matter fluid ruled by $S_m$ and
the function $\mathcal{P}$ is defined in terms of $\Gamma$
and $\Delta$ in (\ref{gamdel}) as
\begin{equation}
\label{bigp}
\mathcal{P}=\sqrt{\frac{1+\Delta}{1-\Gamma}}\cdot
\end{equation}
The careful reader should have noticed that, after varying over $N$,
we set the gravitational gauge to $N=1$, meaning that we work with the
synchronous time coordinate (another convenient choice for the study
of the gauge dynamics in the Yang-Mills regime is the conformal gauge
$N=a$ as it naturally exhibits the conformal invariance).
The Friedmann equation allows us to define the Born-Infeld effective energy
density of the
gauge field as a generalisation of what was proposed in
\cite{dyadichev}
\begin{equation}
\label{rhobi}
\rho_{BI}=\epsilon_{c}A^4(\phi)\left(\mathcal{P}-1\right).
\end{equation}
The Euler-Lagrange equation for the scale factor $a$
gives the acceleration equation:
\begin{equation}
\label{acc}
\frac{\ddot{a}}{a}=\frac{\kappa}{3}\left[\left(V(\phi)-\dot{\phi}^2\right)+\epsilon_{c}A^4(\phi)\left(\mathcal{P}^{-1}-1\right)-\frac{1}{2}\left(\rho_m+3p_m\right)\right],
\end{equation}
where $p_m$ stands for the pressure of the additional matter fluid.
This allow us to define the Born-Infeld effective pressure
\begin{equation}
p_{BI}=\frac{\epsilon_{c}}{3}A^4(\phi)\left(3-\mathcal{P}-2\mathcal{P}^{-1}\right)
\end{equation}
and the equation of state
\begin{equation}
\label{eos}
\lambda_{BI}=\frac{p_{BI}}{\rho_{BI}}=\frac{1}{3}\left(\frac{\epsilon_{c}A^4(\phi)-\rho_{BI}}{\epsilon_{c}A^4(\phi)+\rho_{BI}}\right)
\end{equation}
for the gauge part of the EBID system as in
\cite{dyadichev}. Here we see that the $su(2)-$valued gauge fields
ruled by Born-Infeld lagrangian can be represented by a fluid with an
equation of state that varies continuously from $-\frac{1}{3}$, when
the BI energy density is much larger that the 'critical field'
$A^{-4}(\phi)\rho_{BI}\gg\epsilon_{c}$, to $\frac{1}{3}$ at low energies
$A^{-4}(\phi)\rho_{BI}\ll\epsilon_{c}$.
These two extreme regimes correspond to a gas
of Nambu-Goto strings in three spatial dimensions on one hand (strong
field limit) and radiations on the other (weak field
limit). The transition between these regimes occurs at vanishing pressure
when the BI energy
density is of order of the BI critical energy scale
$\epsilon_{c}$. At low energies
($\epsilon_{c}\rightarrow \infty$), the gauge field
behaves like radiation as expected because the BI lagrangian reduces
to the conformally invariant Yang-Mills one. \\
\\
It is also important to notice that there is no cosmic acceleration
with the metric $g_{\mu\nu}$ as long as the dilaton is massless.
Indeed, the highest value of $\ddot{a}$ that can be
achieved in this frame is identically zero (see \ref{acc}),
in the limit of the pure Einstein-Born-Infeld system
at high energies ($\dot{\phi}=0$, $\lambda_{BI}=-\frac{1}{3}$). However,
we will see that cosmic acceleration may appear once we examine the behaviour
in another frame.
\\
\\
By varying the action (\ref{ebid}) with respect to the dilaton $\phi$,
we find the Klein-Gordon equation:
\begin{eqnarray}
\label{kg}
\ddot{\phi}+3\frac{\dot{a}}{a}\dot{\phi}&=&-\frac{dV(\phi)}{d\phi}-2\epsilon_{c} A^4(\phi)\left[\beta(\phi)\left(\mathcal{P}+\mathcal{P}^{-1}-2\mathcal{R}\right)+ \alpha(\phi) \left(2\mathcal{R}-2\right)\right]\nonumber\\
&& -\gamma(\phi)\left(\rho_m-3p_m\right)
\end{eqnarray}
where $\alpha(\phi)= \frac{d\ln A(\phi)}{d\phi}$, $\beta(\phi)= \frac{d\ln B(\phi)}{d\phi}$ and $\gamma(\phi)= \frac{d\ln C(\phi)}{d\phi}$.
The key point of the physics in the EBID system lies in the fact that
the dilaton field couples differently to the gauge sector of the
theory (last term) depending on the values of the coupling functions $A$ and
$B$. Although the cosmological dynamics of the gauge fields ruled
by Born-Infeld lagrangian can be regarded as a fluid with an equation
of state given by (\ref{eos}), the coupled dynamics of the dilaton and the
gauge field does not reduce in general
to a scalar-tensor theory with this
fluid as background. This is only the case when we have a universal coupling
to the metric $g_{\mu\nu}$, i.e. when $A=B$.
In the general case, there exists a non-trivial energy exchange
between the dilaton and the gauge sectors of the theory that will
dominate at late epochs as we shall see further. \\
\\
Finally, the Euler-Lagrange equation for the gauge field $\sigma$
gives the Born-Infeld equation that rules the gauge potential dynamics:
\begin{equation}
\label{bi}
\ddot{\sigma}+2\frac{\sigma^3}{a^2}\mathcal{P}^{-2}-2\frac{\dot{a}}{a}\dot{\sigma}\left(\frac{1}{2}-\mathcal{P}^{-2}\right)
+2\dot{\phi}\dot{\sigma}\left[2\alpha(\phi)\frac{\mathcal{R}}{\mathcal{P}}-\beta(\phi)\left(2\frac{\mathcal{R}}{\mathcal{P}}+1-\mathcal{P}^{-2}\right)\right]=0\cdot
\end{equation}
This equation is essentially the same that in \cite{dyadichev} except
from the coupling term proportional to $\dot{\phi}$
which accounts for the direct energy exchange at the microscopic level between
the fields. It is important to notice that we did not assume any direct coupling between the gauge field
and the additional matter fluid, which will allow to treat them separately.\\
\\
Indeed, from equation (\ref{bi}), and following
\cite{dyadichev}, it is possible
to derive an energy conservation equation for the BI density :
\begin{eqnarray}
\label{cons}
\dot{\rho_{BI}}&=&-2\frac{\dot{a}}{a}\rho_{BI}\frac{\rho_{BI}+2\epsilon_{c}A^4(\phi)}{\rho_{BI}+\epsilon_{c}A^4(\phi)}+4\alpha(\phi)\dot{\phi}\epsilon_{c}A^4(\phi)\left(\mathcal{R}-1\right)\nonumber\\
&&-2\beta(\phi)\dot{\phi}\epsilon_{c}A^4(\phi)\left(\frac{\Delta+\Gamma}{\mathcal{R}}-2\mathcal{P}+2\mathcal{R}\right)\cdot
\end{eqnarray}
Now that we have derived the complete set of the EBID field equations
for cosmology,
we propose the reader to briefly review some basic features of
Born-Infeld cosmology on one hand and dilaton cosmology on the
other. In the rest of this paper, we will only consider a massless
dilaton (i.e., vanishing
self-interaction potential $V(\phi)=0$).
\section{Born-Infeld and Dilaton Cosmologies}
\subsection{Non-Abelian Born-Infeld Cosmology}
The Non-Abelian Born-Infeld cosmology in various spacetimes with different
values of the curvature and the cosmological constant
was described in detail in \cite{dyadichev}. The field equations governing these models are those of the previous section
with a vanishing dilaton $\phi=\dot{\phi}=0$, constant coupling functions
$A=B=1$ and no additional matter fluid $\rho_m=0$.
The equation (\ref{cons}) for BI energy conservation can be written as
\begin{equation}
\label{consbis}
\dot{\rho_{BI}}=-2\frac{\dot{a}}{a}\rho_{BI}\frac{\rho_{BI}+2\epsilon_{c}}{\rho_{BI}+\epsilon_{c}}
\end{equation}
which admits a first integral:
\begin{equation}
\label{fi}
a^4\rho_{BI}\left(\rho_{BI}+2\epsilon_{c}\right)=\mathcal{C}
\end{equation}
where $\mathcal{C}$ is a positive constant. In the strong field limit, $\rho_{BI}\gg\epsilon_{c}$,
the BI energy density redshifts as $\rho_{BI}\approx a^{-2}$ while in the weak limit, $\rho_{BI}\ll\epsilon_{c}$,
we retrieve the radiation behaviour $\rho_{BI}\approx a^{-4}$ characteristic of the conformal invariance of the gauge field at such energies.
This allows to treat separately the spacetime evolution and the dynamics of the gauge field.
Although the complete analytical solutions for both gravitational and gauge sector were derived in \cite{dyadichev}, let us
illustrate simply the main features of this cosmological model.\\
\\
First, the strong field limit $\rho_{BI}\gg\epsilon_{c}$ corresponds
to $\mathcal{P}\gg 1$. In this limit, the acceleration equation
(\ref{acc}) reduces to
$$
\frac{\ddot{a}}{a}=-\frac{\kappa}{3}\epsilon_{c}
$$
whose general solution is
$$
a(t)=a^*\sin\left(\sqrt{\frac{\kappa\epsilon_c}{3}} t\right)
$$
where we set $a(0)=0$ and $a^*$ is the value of the scale factor at
the time $t^*=\sqrt{\frac{3}{\kappa\epsilon_c}}\frac{\pi}{2}$ (in
Planck units). Therefore, the cosmic expansion starts with a zero
acceleration at the singularity. Then, setting $\mathcal{P}\gg 1$ in
equation (\ref{bi}) brings
$$
\ddot{\sigma}-\frac{\dot{a}}{a}\dot{\sigma}=0
$$
which shows that $\dot{\sigma}$ scales as $a$. Near the singularity, the behaviour of the gauge field is therefore
$$
\sigma(t)=\mp \frac{\sqrt{3\epsilon_c}}{3} a^*\cos\left((\sqrt{\frac{\kappa\epsilon_c}{3} t}\right),
$$
and the gauge potential $\sigma$ starts at rest.\\
\\
Then, in the weak field regime, $\rho_{BI}\ll\epsilon_{c}$ and
$\mathcal{P}\approx 1$ ($\epsilon_c\rightarrow\infty$).
This limit corresponds to the Einstein-Yang-Mills cosmological
solution studied in \cite{galtsov}. The conformal invariance of the
gauge field in that regime yields that the scale factor behaves like
in the radiation-dominated era : $a(t)\approx \sqrt{t}$ (in
synhronous time). On the other hand, the energy conservation
equation (\ref{cons}) now reduces to
$$
a^2\dot{\sigma}^2+\sigma^4=\frac{\mathcal{C}}{3\epsilon_{c}}
$$
which can be integrated in terms of the Jacobi elliptic function. Moving to the conformal time coordinate $dt=ad\eta$,
we find
$$
\sigma(\eta)=\mathcal{E}^{1/4}cn\left(\mathcal{E}^{1/4}\eta;-1\right),
$$
where $cn(u,k)$ is the Jacobi elliptic function and $\mathcal{E}=\frac{\mathcal{C}}{3\epsilon_{c}}$. In synchronous time, the gauge potential $\sigma$ oscillates
with a fixed amplitude and a growing period.\\
\\
More generally, it is possible to derive a general solution for the
gauge potential. Let us rewrite the first integral (\ref{fi}) in
terms of $\mathcal{P} $ as
\begin{equation}
\label{bigp_bi} \mathcal{P}=\sqrt{1+\frac{\mathcal{C}}{\epsilon_c^2
a^4}}\cdot
\end{equation}

\begin{figure}
\begin{center}
\begin{tabular}{c}
\includegraphics[scale=0.4]{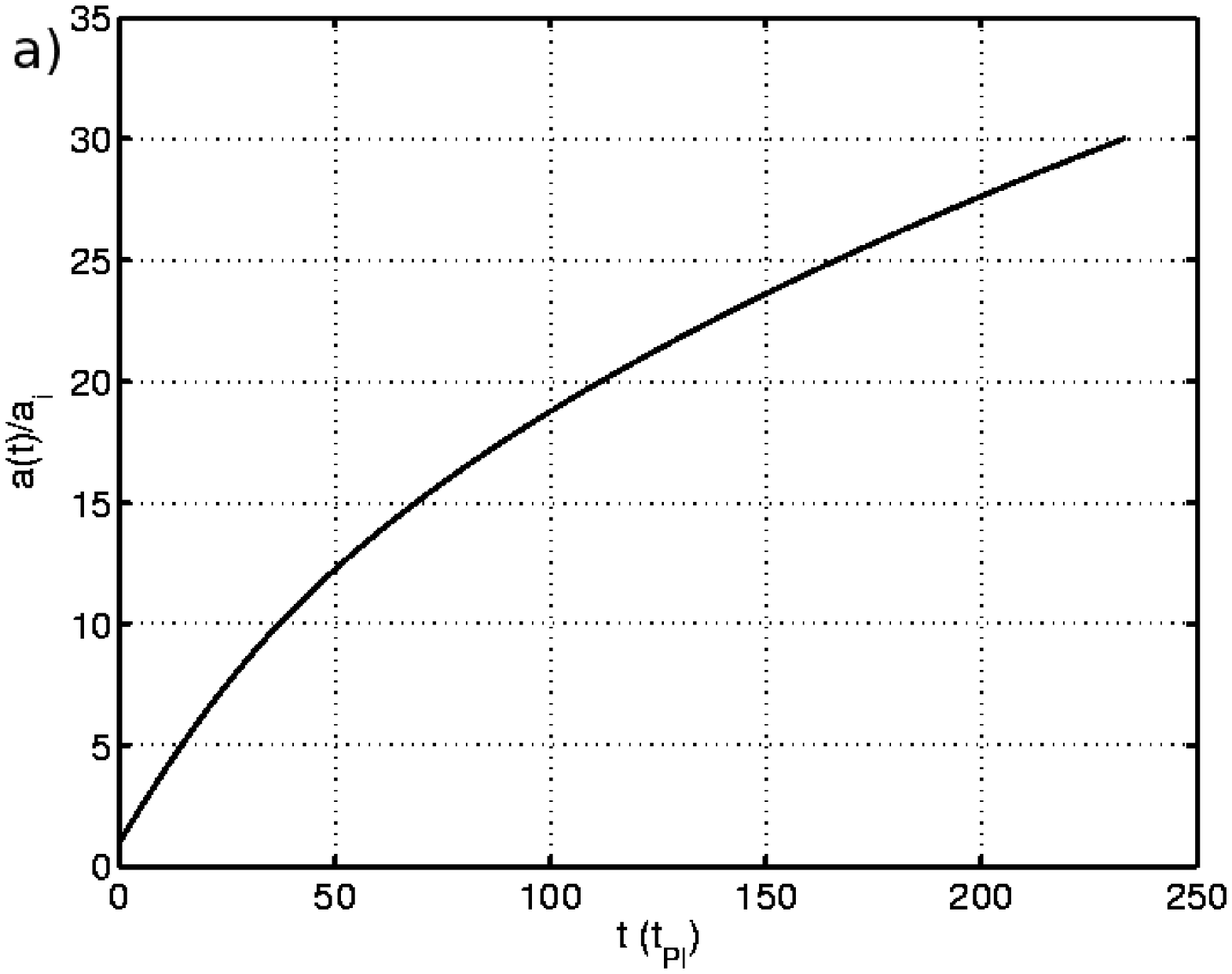}
\includegraphics[scale=0.4]{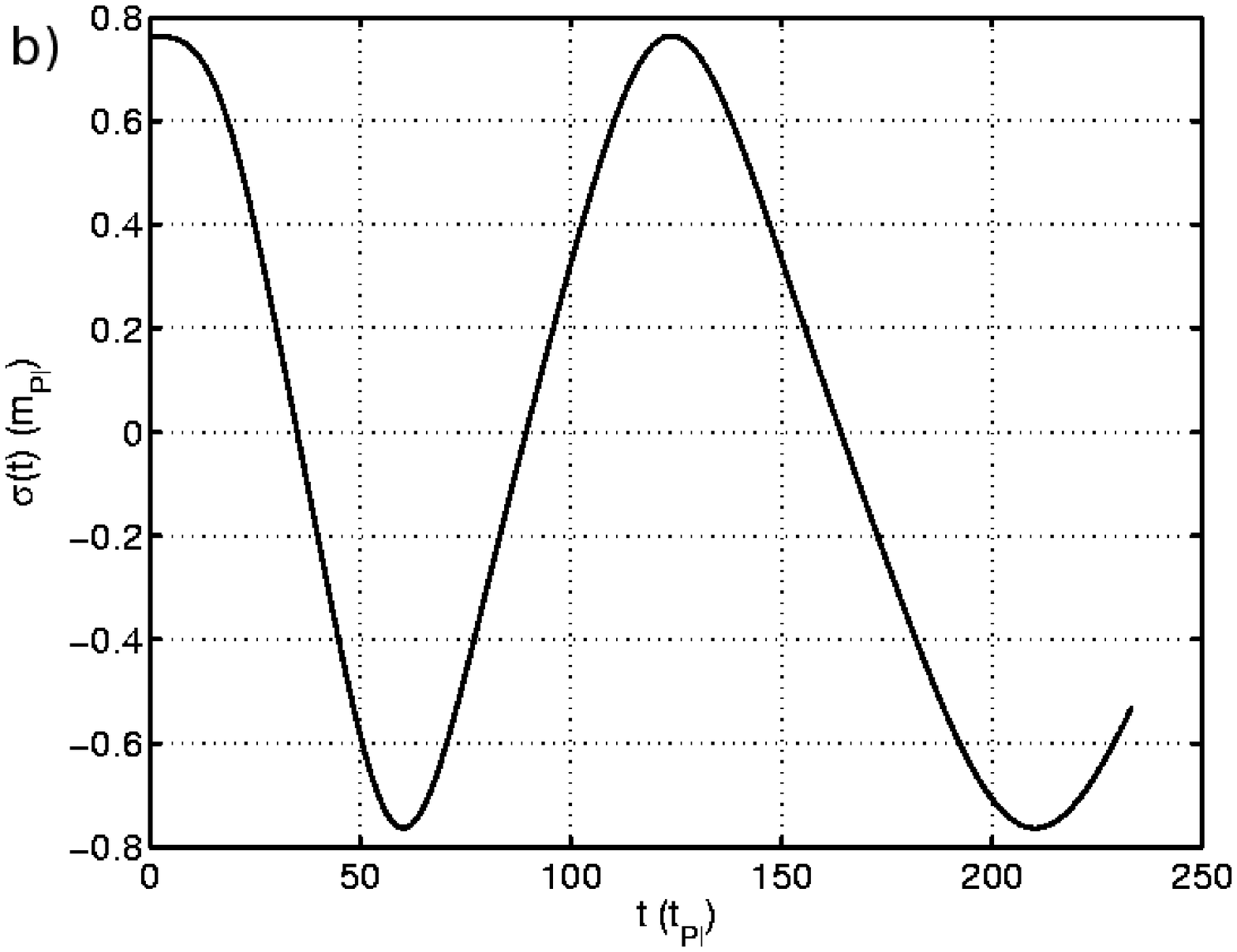}
\includegraphics[scale=0.4]{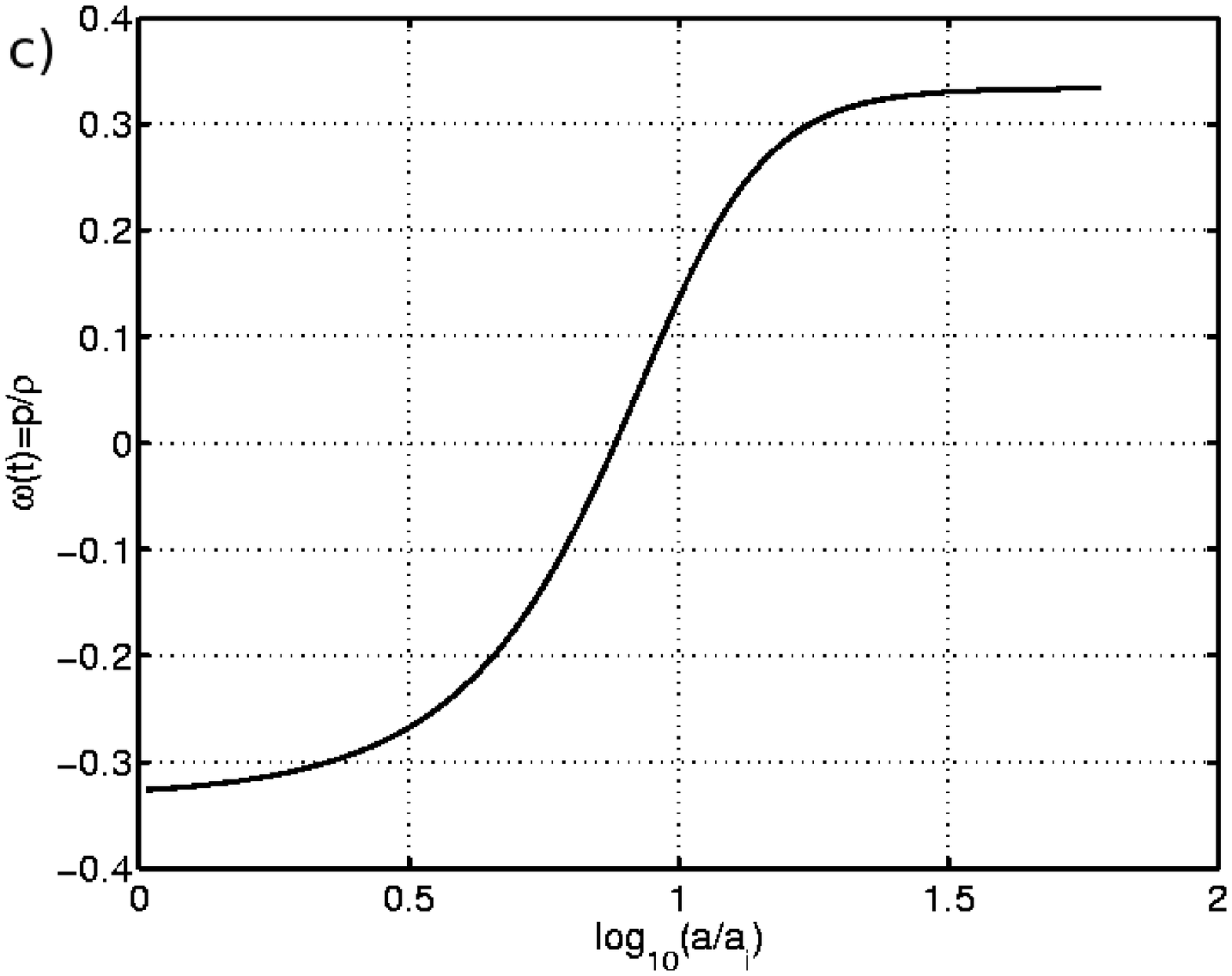}
\end{tabular}
\end{center}
\caption{Illustration of Non-Abelian Born-Infeld cosmology :
a) scale factor
b) gauge potential
c) equation of state of the Born-Infeld ``\textit{fluid}''}
\label{bicos}
\end{figure}

Using the definitions (\ref{bigp}) and (\ref{gamdel}), the previous equation may be integrated to give the gauge potential
(in the conformal gauge $dt=ad\eta$):
$$
\sigma(\eta)=a_0^2\sqrt{\epsilon_c} cn\left(\mathcal{P}^{-1}\eta;-1\right)\cdot
$$
Figure \ref{bicos} illustrates the evolution of the
scale factor, the gauge potential and the equation of state during the
expansion of a non-abelian Born-Infeld universe.
The figures correspond to the numerical integration of equations
(\ref{acc}) and (\ref{bi}) with $\phi=\dot{\phi}=0$ and $A=B=1$.
During numerical evolution, we monitor
the violation of the hamiltonian constraint (\ref{hebid}) (see the appendix
for more details on integrating the EBID system). In this
case, this violation does not exceed a part on $10^{-12}$. Initial
conditions at $t_i=0$ were assumed such as $\rho_{BI}(t_i)=100
\epsilon_{c}$, $a(t_i)=a_i=1$ and $\dot{\sigma}(t_i)=0$ ($\epsilon_c=10^{-4}\times m_{Pl}^4$).
\subsection{Dilaton Cosmology}
Dilaton cosmology can retrieved from our fundamental equations by setting $\sigma$ equal to $0$.
Indeed, tensor-scalar theories can be written in the so-called ``\textit{Einstein}''
conformal frame:
\begin{equation}
\label{einstein}
S=\frac{1}{2\kappa}\int d^4 x
\sqrt{-g_*}\left\{R_*-2g_*^{\mu\nu}\partial_\mu\varphi\partial_\nu\varphi\right\}+S_m\left[\psi_m,C^2(\varphi)g^*_{\mu\nu}\right],
\end{equation}
where $\kappa$ is therefore the ``\textit{bare}'' gravitational
coupling constant, $S_m$ is the action for the
matter fields $\psi_m$, $\varphi=\sqrt{\kappa/2}\phi$ and $g^*_{\mu\nu}$ the
``\textit{Einstein}'' metric tensor which corresponds to basic gravitational
variables with pure spin 2 propagation modes. This metric is measured
by using purely gravitational rods and clocks and allows to account
for the dynamics in a simpler way\footnote{In particular, this frame
represents gravitation with its pure scalar and tensor degrees of freedom
and the limit of general relativity is not singular.} than a observable frame
in which the metric tensor $\tilde{g}_{\mu\nu}$ is universally
coupled to matter fields $\psi_m$.
This frame is called the ``\textit{Jordan-Fierz}''
frame in which the action (\ref{einstein}) can be written
\begin{equation}
\label{jordan}
S=\frac{1}{2}\int d^4 x
\sqrt{-\tilde{g}}\left\{\Phi
  \tilde{R}-\frac{\omega\left(\Phi\right)}{\Phi}\tilde{g}^{\mu\nu}\partial_\mu\Phi\partial_\nu\Phi\right\}+S_m\left[\psi_m,\tilde{g}_{\mu\nu}\right],
\end{equation}
where $\tilde{R}$ is the curvature scalar build upon the physical
metric $\tilde{g}_{\mu\nu}$ which is measured using
non-gravitational rods and clocks and where
$\omega\left(\Phi\right)$ is called the coupling function. The
scalar field $\Phi$ now gives the effective gravitational coupling
constant. In this frame, matter fields evolves in the same way that
they could do in general relativity because the action of matter
does not depend explicitly on the scalar field $\phi$, as matter
couples universally to the physical metric $\tilde{g}_{\mu\nu}$.
Einstein and Jordan frames can be linked together through the
conformal transformation
\begin{equation}
\tilde{g}_{\mu\nu}=C^2\left(\varphi\right)g^*_{\mu\nu}
\end{equation}
and the following relation between the scalar field $\phi$
 in the Einstein frame (pure spin 0 dynamics) and
its counterpart $\Phi$ in the Jordan physical frame:
\begin{equation}
\Phi^{-1}=\kappa C^2\left(\varphi\right),
\end{equation}
with $\varphi=\sqrt{\kappa/2}\phi$.
The energy density and pressure of the matter fluid in both frames are related by
\begin{eqnarray}
\rho_m&=&C^4\tilde{\rho_m}\label{jordan2}\\
p_m&=&C^4\tilde{p_m},
\end{eqnarray}
where $\rho_m$ and $p_m$ represent these quantities expressed in the Einstein frame.
At the opposite of scalar-tensor theories, we did not assume here an universal coupling to the metric tensor $g_{\mu\nu}$.
Indeed, the action (\ref{lebid}) reduces to a tensor-scalar theory in the Einstein frame (\ref{einstein}) only
when $A(\phi)=B(\phi)=C(\phi)$, i.e. when the weak equivalence principle applies.
Let us now remind the reader about the behaviour of such tensor-scalar
theories in the presence of a background cosmological fluid.
When the matter fields $\psi_m$ are represented
under the approximation of a perfect fluid,
the dynamics
of the scalar field
is ruled by (see \cite{damour,alimi,barrow,serna}):
\begin{equation}
\label{ts}
\frac{2}{\left(3-\varphi^{'2}\right)}\varphi''+\left(1-\lambda\right)\varphi'+(1-3\lambda)\gamma(\varphi)=0,
\end{equation}
where $\gamma(\varphi)=\frac{d\ln C(\varphi)}{d\varphi}$ and
$\lambda=\rho/p$ is the equation of state for the cosmological
fluid. In the previous
equation, a prime denotes the derivative with respect to the variable $p=\ln(a/a_i)$.
The action of the cosmological fluid is thus to damp the dynamics of the
scalar field while it is rolling down some effective potential depending on the
coupling function.
Furthermore, the scalar field now has an effective, velocity-dependent, mass of
\begin{equation}
m(\varphi)=\frac{2}{\left(3-\varphi^{'2}\right)}
\end{equation}
where the field has a limiting speed $\varphi'\le \sqrt{3}$ for which its effective
mass diverges. This relativistic limit corresponds to the case where
the energy density of the background fluid is negligible compared
to the kinetic energy of the scalar field (the universe is dominated by the kinetic energy of the scalar field).
\subsection{The universal coupling for EBID cosmology}
Let us now turn back to the EBID system we wrote in the previous section and focus on the gauge sector only by setting
$S_m=0$. If we now assume
a universal coupling to the metric $g_{\mu\nu}$ by setting $A=B$, we now have for
the dilaton equation (\ref{kg}):
\begin{equation}
\ddot{\phi}+3\frac{\dot{a}}{a}\dot{\phi}+\alpha(\phi)\left(\rho_{BI}-3p_{BI}\right)=0,
\end{equation}
where
$$
\rho_{BI}-3p_{BI}=2\epsilon_c A^4(\phi)\left(\mathcal{P}+\mathcal{P}^{-1}-2\right).
$$
In this case of universal coupling, we recognize the equation for the scalar field in presence of
a background fluid for general tensor-scalar theories\footnote{We also have $\rho_{BI}=A^4 \tilde{\rho_{BI}}$
(see \ref{rhobi}) for the relation between the energy density expressed in the Einstein and Jordan frames (quantities with a $\tilde{\cdot}$).}.
The equations (\ref{hebid}), (\ref{acc}) and (\ref{kg})
can be simply solved in the Jordan frame by using the results on the non-abelian Born-Infeld cosmology (previous paragraph).
The equation (\ref{bi}) for the gauge field in the Einstein frame can also be written
\begin{equation}
\ddot{\sigma}+2\frac{\sigma^3}{a^2}\mathcal{P}^{-2}-2\frac{\dot{a}}{a}\dot{\sigma}\left(\frac{1}{2}-\mathcal{P}^{-2}\right)-2\alpha(\phi)\dot{\phi}\dot{\sigma}\left(1-\mathcal{P}^{-2}\right)=0\cdot
\end{equation}
Therefore, in the weak energy regime $\rho_{BI}\ll\epsilon_c$
($\mathcal{P}\approx 1$), the gauge field undergoes a conformally
invariant dynamics (non-abelian radiation, $\rho_{BI}=3p_{BI}$) and
decouples from the scalar field.

In a radiation-dominated universe, where the equation of state is
$\lambda=p/\rho=1/3$, the dynamics of the scalar field is
given by the following solution to (\ref{ts}) (cf. \cite{damour,alimi,serna}):
\begin{equation}
\label{phirad}
\varphi(p)=\varphi_\infty-\sqrt{3}\ln\left[Ke^{-p}+\sqrt{1+K^2e^{-2p}}\right],
\end{equation}
where the integration constant $K$ is determined from the initial
velocity $\varphi'(p=0)=\varphi'_0$:
$$
K=\frac{\varphi'_0}{3-\varphi^{'2}_0}\cdot
$$
This should correspond to the low-energy limit of the Born-Infeld field equations
when a universal coupling is assumed: the scalar field velocity in p-time should be damped to zero by the cosmological expansion.
It should be noticed that when there is
no universal coupling $A\ne B$, we do keep an energy exchange between the dilaton and the gauge fields and the usual dynamics
of tensor-scalar theories will be modified.
Moreover, when there is no universal coupling, the energy exchange between the gauge
potentials and the dilaton field will prevent the dynamics to be
purely dictated by the solution (\ref{phirad}). As we shall see in
section 5, this solution will accurately describe the early epochs of
evolution when the field is almost relativistic. However, at late
times, energy transfer between dilaton and gauge fields will
substantially alter the dynamics. \\
\\
Once again, let us assume a universal coupling and consider the strong field limit of the Born-Infeld system where we have $\lambda=-1/3$.
The dilaton equation (\ref{ts}) now becomes
\begin{equation}
\label{ts_string}
\frac{\varphi''}{3-\varphi^{'2}}+\frac{2}{3}\varphi'+\alpha(\varphi)=0\cdot
\end{equation}
Let us now write down for the dilaton coupling function
\begin{eqnarray}
A^2\left(\phi\right)&=&e^{k\phi} \label{conformal} \\
\omega(\Phi)&=&\frac{2\kappa-3k^2}{2 k^2}\\
\alpha(\varphi)&=&\frac{k}{\sqrt{2\kappa}}\\
|3+2\omega(\Phi)|&=&\alpha^{-2}(\varphi)
\end{eqnarray}
and face the simplest tensor-scalar theory of Brans-Dicke type ($\varphi=\sqrt{\kappa/2}\phi$).
Using CONVODE \cite{convode}, it is possible to find an analytic
solution for $\varphi'$ under the following implicit form when we use the coupling function (\ref{conformal}):
\begin{equation}
\label{sol_string}
\left(6\mathcal{A}^2-8\right)(p+p_0)=\sqrt{3}\mathcal{A}\ln\left(\left|\frac{\varphi'-\sqrt{3}}{\varphi'+\sqrt{3}}\right|\right)-2\ln\left(\frac{\varphi^{'2}-3}{\left(3\mathcal{A}+2\varphi'\right)^2}\right)
\end{equation}
where $\mathcal{A}=k/\sqrt{2\kappa}$ and $p_0$ some integration constant. When $p\rightarrow\infty$,
there is an attractor for $\varphi'$, namely
\begin{equation}
\label{att_string}
\varphi'(p\rightarrow\infty)=-\frac{3}{2}\frac{k}{\sqrt{2\kappa}}\cdot
\end{equation}
Therefore, there is also a maximum value for the dilatonic coupling
constant $k$ for which the attractor corresponds to the relativistic
limit for the dilaton ($|\varphi'_\infty|\rightarrow\sqrt{3}$):
\begin{equation}
\label{kmax_string}
k_{max}=\sqrt{\frac{8\kappa}{3}}\cdot
\end{equation}
In the non-relativistic limit $\left|\varphi'\right|\ll \sqrt{3}$,
equation (\ref{ts_string}) now becomes
$$
\varphi''+2\varphi'+3\mathcal{A}=0,
$$
which can be solved easily to give
\begin{equation}
\label{nr_phibi}
\varphi(p)=-\frac{3}{2}\frac{k}{\sqrt{2\kappa}}\left(p+\frac{1}{2}e^{-2p}-\frac{1}{2}\right),
\end{equation}
where we assumed $\varphi_i=0$.
We can see that, due to the constant potential term in equation
(\ref{ts_string}), the value of the dilaton field goes to $-\infty$
($+\infty$ if $k<0$) with the time-variable $p$. However, the
gauge energy density will also decrease with time and finally the
assumption of strong field will be no longer true as
$\rho_{BI}$ becomes less than the critical energy $\epsilon_{c}$. At the end of the evolution, we should retrieve
the radiation case for which the solution in case of universal coupling was described above.
Once again, a non-universal coupling to the metric yields modification of these behaviours as we shall see further.
\\
\\
Now that we have recalled the main features of Born-Infeld and dilaton
cosmologies as well as EBID system with universal coupling, let us now discuss how the non universal coupling in the general EBID system
will modify a tensor-scalar cosmological
picture. This will be done in three steps: in the following section,
we will focus on the strong field limit where the gauge field mimics a
Nambu-Goto string gas ; then on the low-energy limit which corresponds
to the Yang-Mills regime for the gauge fields and finally to the
general cosmological evolution where transition between both regimes occurs.
\section{The Strong Field Regime}
\begin{figure}
\begin{center}
\begin{tabular}{cc}
\includegraphics[scale=0.3]{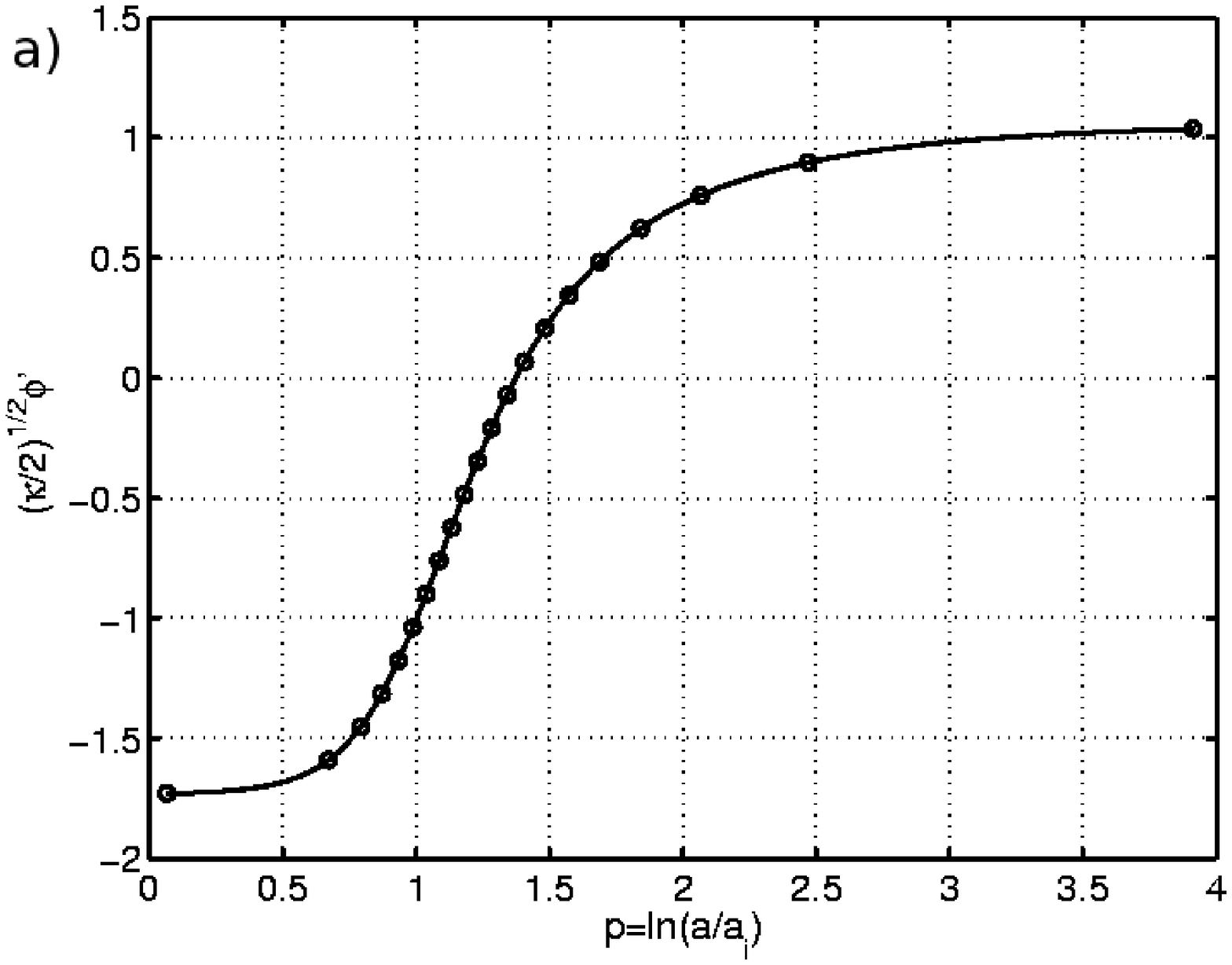} &
\includegraphics[scale=0.3]{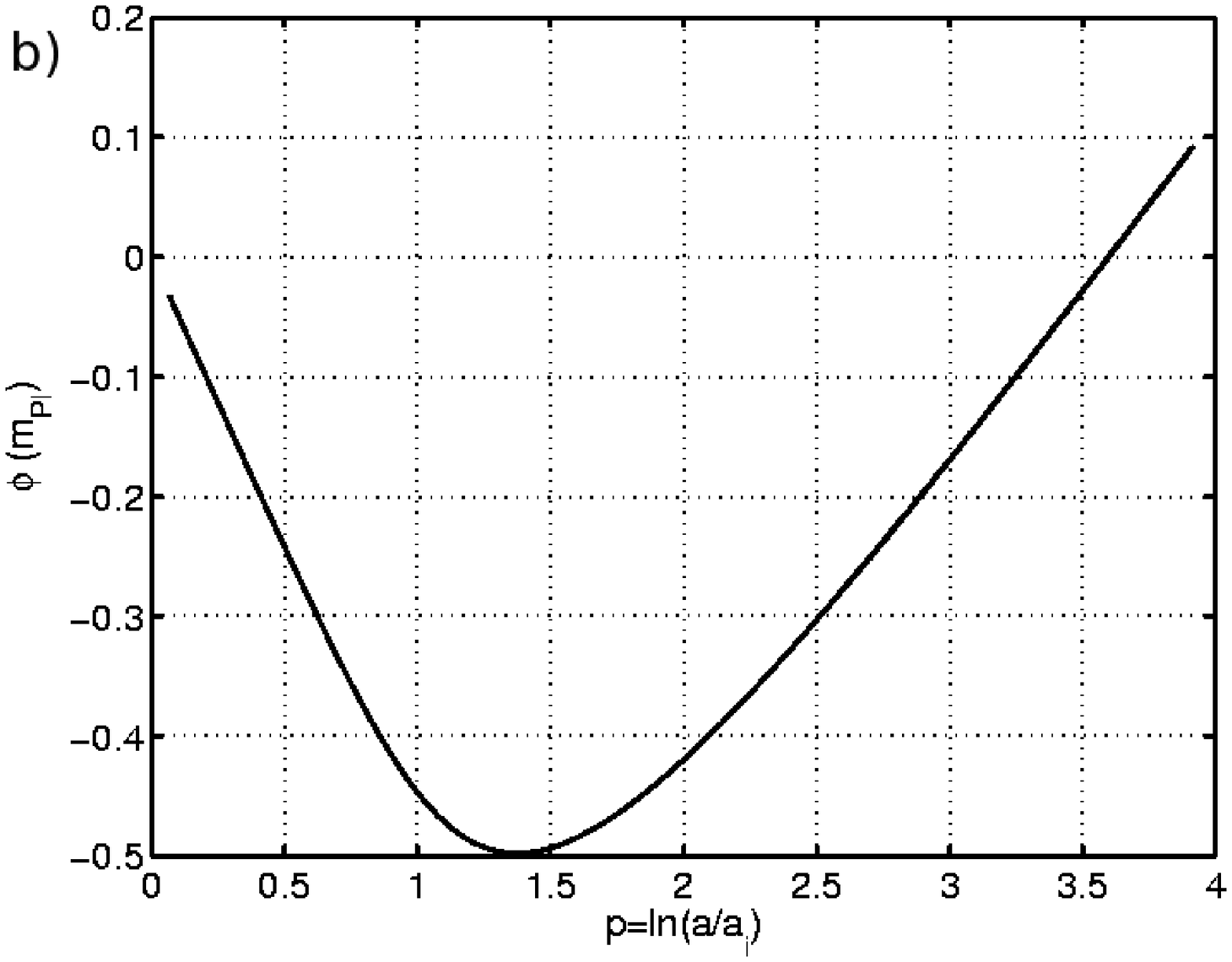}\\
\includegraphics[scale=0.3]{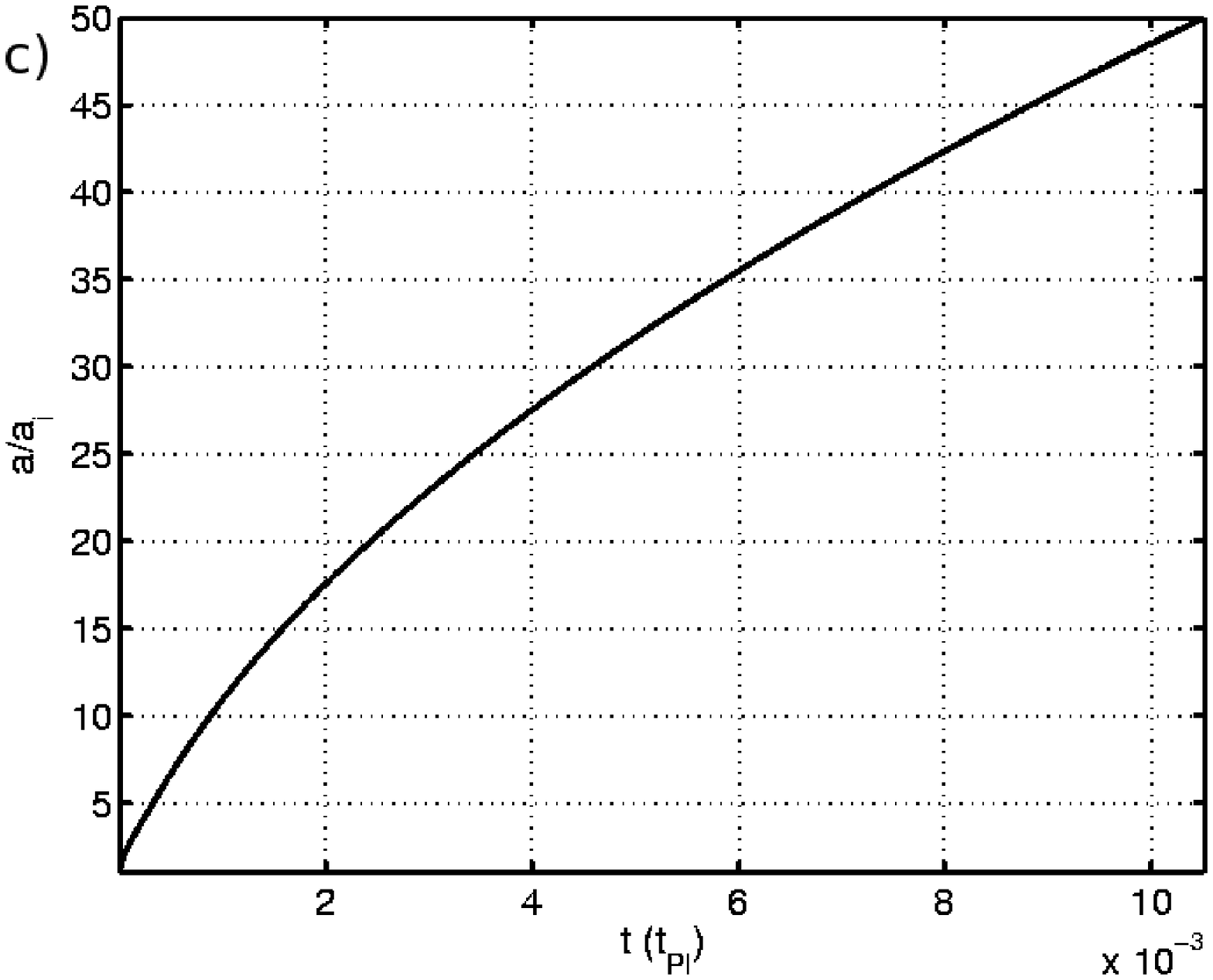} &
\includegraphics[scale=0.3]{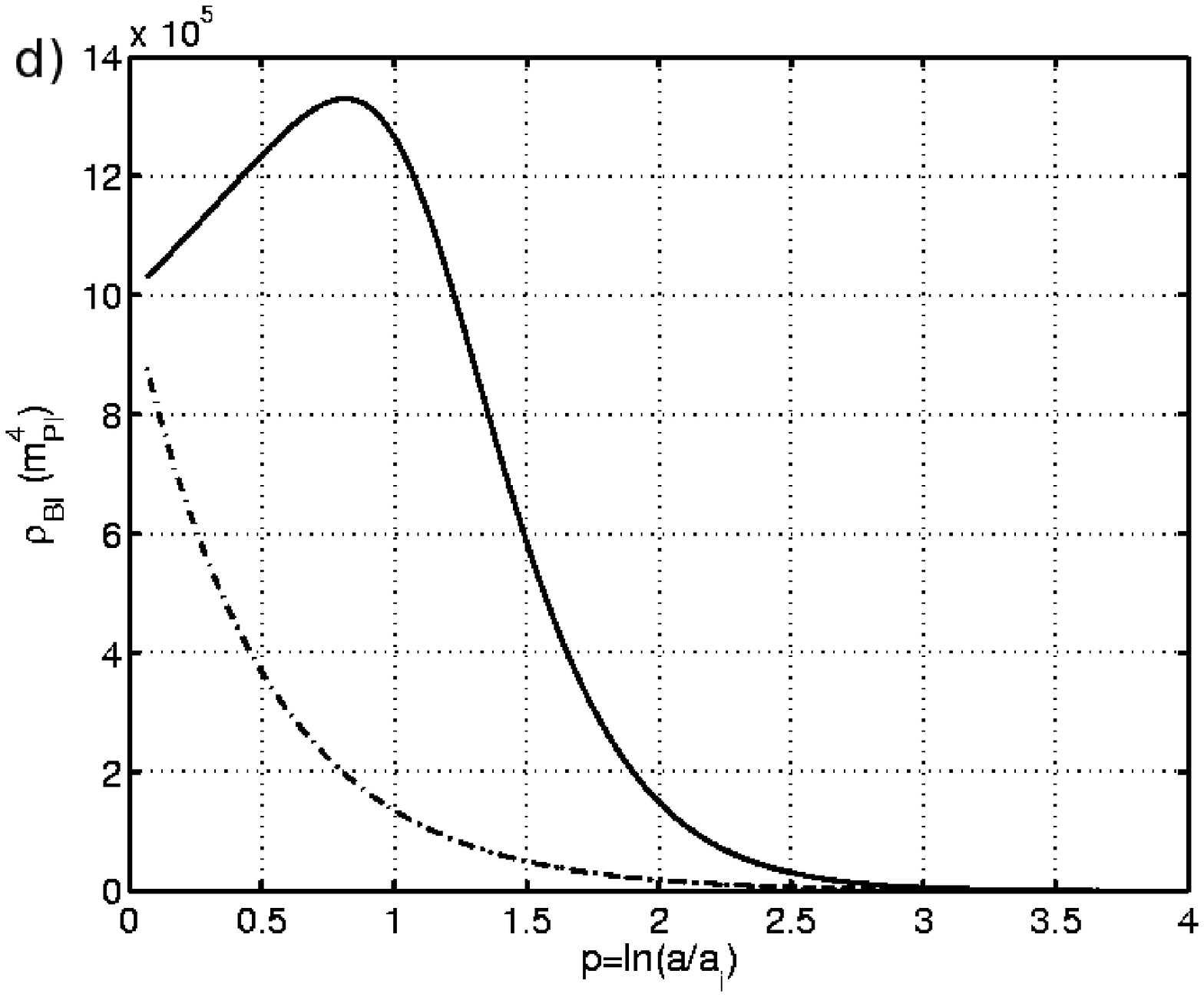}
\end{tabular}
\end{center}
\caption{Evolution of the EBID system in the strong-field regime with the p-time variable
a) $\varphi'$
b) dilaton field $\phi$
c) scale factor $a(t)$
d) Energy density of the BI gauge field
($a_i=1$, $\rho_{BI}(a_i)/\epsilon_{c}=10^{10}$, $\epsilon_c=10^{-4}\times
m_{Pl}^4$, $k=5$, $\phi'_i=-1.73$, $\delta H/H<10^{-13}$)}
\label{st_regime}
\end{figure}

As we have seen earlier, the strong field limit is reached when the
BI critical energy $\epsilon_c$ can be neglected with regards to the
gauge field energy density. Setting $\mathcal{P}\gg 1$ into the EBID
field equations (\ref{hebid}), (\ref{acc}) and (\ref{kg}) and
(\ref{bi}) with $\rho_m=p_m=0$ and $V=0$, we find:
\begin{eqnarray}
\left(\frac{\dot{a}}{a}\right)^2&=&\frac{\kappa}{3}\left[\frac{\dot{\phi}^2}{2}+\epsilon_{c}A^4(\phi)\mathcal{P}\right]\label{sf_h}\\
\frac{\ddot{a}}{a}&=&-\frac{\kappa}{3}\left[\dot{\phi}^2+\epsilon_{c}A^4(\phi)\right]\label{sf_acc}\\
\ddot{\phi}&+&3\frac{\dot{a}}{a}\dot{\phi}+2\epsilon_{c} A^4(\phi)\mathcal{P}\left[-\beta(\phi)+ 2 \alpha(\phi) \right]=0
\label{sf_kg}\\
\ddot{\sigma}&-&\frac{\dot{a}}{a}\dot{\sigma}+2\dot{\phi}\dot{\sigma}\left[2\alpha(\phi)-3\beta(\phi)\right]=0\cdot
\label{sf_bi}
\end{eqnarray}
where we used $\frac{\mathcal{R}}{\mathcal{P}}=1-\Gamma$ and assumed
$\Gamma\ll 1$ which will be verified afterwards by the agreement
between analytical and numerical solutions.
\\
\\
In the following, we will assume the exponential coupling function (\ref{conformal})
for the sake of simplicity. However, the qualitative analysis will be valid for any coupling functions.
\\
\\
In terms of the new time variable
$$
p=\ln\left(\frac{a}{a_i}\right)
$$
(where $a_i$ defines the initial zero value for $p$) we can combine
equations (\ref{sf_h}) to (\ref{sf_kg}) to obtain
\begin{equation}
\label{sf_phi}
\frac{\varphi''}{3-\varphi^{'2}}+\frac{2}{3}\varphi'+\left(-\beta(\varphi)+2\alpha(\varphi)\right)=0, 
\end{equation}
where $\varphi=\sqrt{\kappa/2}\phi\cdot$
In order to particularise, we can set now
$\alpha=0$ ($A=1$) and $B(\varphi)=\exp(k/\sqrt{2\kappa}\varphi)$ and find for the dilaton equation
\begin{equation}
\frac{\varphi''}{3-\varphi^{'2}}+\frac{2}{3}\varphi'-\frac{k}{\sqrt{2\kappa}}=0\nonumber
\end{equation}
which admits a solution similar to the case of a tensor-scalar theory with Nambu-Goto string gas (\ref{ts_string}):
\begin{equation}
\label{sol_bi}
\left(6\mathcal{A}^2-8\right)(p+p_0)=-\sqrt{3}\mathcal{A}\ln\left(\left|\frac{\varphi'-\sqrt{3}}{\varphi'+\sqrt{3}}\right|\right)-2\ln\left(\frac{\varphi^{'2}-3}{\left(3\mathcal{A}-2\varphi'\right)^2}\right),
\end{equation}
where $\mathcal{A}=k/\sqrt{2\kappa}\cdot$ When $p\rightarrow\infty$,
the attractor for $\varphi'$ is now exactly the opposite of the universal coupling case:
\begin{equation}
\label{att_bi}
\varphi'(p\rightarrow\infty)=\frac{3}{2}\frac{k}{\sqrt{2\kappa}}.
\end{equation}
The maximum value for the dilatonic coupling
constant $k$ is the same as before (equation (\ref{kmax_string})) and the non-relativistic limit
can be obtained from (\ref{nr_phibi}) with an opposite sign.
The constant potential term in equation
(\ref{sf_phi}) is negative so that the value of the dilaton field is pushed toward $+\infty$
($-\infty$ if $k<0$) with the time-variable $p$, as long as the gauge field remains in the strong field
limit $\rho_{BI}\gg\epsilon_{c}$.
\\
\\
Figure \ref{st_regime} a) illustrates the evolution of the dilaton velocity with respect
to the time variable $p$ in the strong field limit. The trajectory has
been computed numerically (solid line in Figure \ref{st_regime} a))
from the integration of the full EBID system with the
initial conditions indicated in the caption (with $A=1$ and $B(\varphi)=\exp(k/\sqrt{2\kappa}\varphi)$). Also shown is the
analytical approximation of the strong field limit given by equation
(\ref{sol_bi}) represented by big dots.
\\
\\
Figure \ref{st_regime} b) gives the evolution of the dilaton field along the
cosmic expansion for the case $A=1$ and $B(\varphi)=\exp(k/\sqrt{2\kappa}\varphi)$. Starting with a negative velocity, the dilaton
is damped to a minimum before being accelerated to infinite
values (with $k>0$).
Fortunately, as the BI
energy density will decrease with time, the strong field limit
$\rho_{BI}\gg \epsilon_{c}$ will soon be no longer valid. We shall see
further that, in the Yang-Mills limit, the same coupling to gauge fields
will bring the dilaton to infinitely negative values. Of
course, similar conclusion can be found when $k$ is negative.
Figure \ref{st_regime} c)  gives the behaviour of the scale factor in the case
discussed here. As the dilaton field is relativistic at infinitely-low
times $p\rightarrow -\infty$ and therefore dominates the energy
content of the universe, the expansion starts with an infinite rate,
breaking the ``\textit{renormalisation}'' that was done in simple BI
cosmologies.
Finally, let us focus on the gauge sector of the EBID system. In the
strong field limit, the features of Born-Infeld cosmologies for the
gauge field are conserved: it starts at rest, damped by the cosmic
expansion, before entering the oscillation regime of the weak energy Yang-Mills
limit.
Figure \ref{st_regime} d) illustrates the evolution of $\mathcal{P}$ for the numerical solution
(solid line) with dilaton compared to the evolution in a simple BI
universe with same initial BI energy density (dash-dotted
line).
This holds for the particular coupling $A=1$ and $B(\varphi)=\exp(k/\sqrt{2\kappa}\varphi)$.
If we now transpose $A$ and $B$, it is easy to see from (\ref{sf_phi}) that the attracting value for $\varphi'$ will be twice the value of
the universal coupling (\ref{att_string}). Therefore, when the coupling to the volume form is weaker than the coupling to the
metric\footnote{For example, in the case $A=1$ and $B(\varphi)=\exp(k/\sqrt{2\kappa}\varphi)$ we just discussed.},
the scalar field behaves just the opposite way than in the universal coupling (with $\lambda=-1/3$).
Up to this point, we have obtained both analytical and numerical
solutions for the strong field limit, $\rho_{BI}\gg\epsilon_{c}$, of the EBID system. We have
also explained qualitatively the effects of non-universal coupling to the Einstein metric $g_{\mu\nu}$
on the dynamics of the scalar field.
Let us now turn to
the weak field regime in which the gauge sector is ruled by Yang-Mills lagrangian.
\section{The Weak Field Regime: solutions of the
  Einstein-Yang-Mills-Dilaton system}
\begin{figure}
\begin{center}
\includegraphics[scale=0.45]{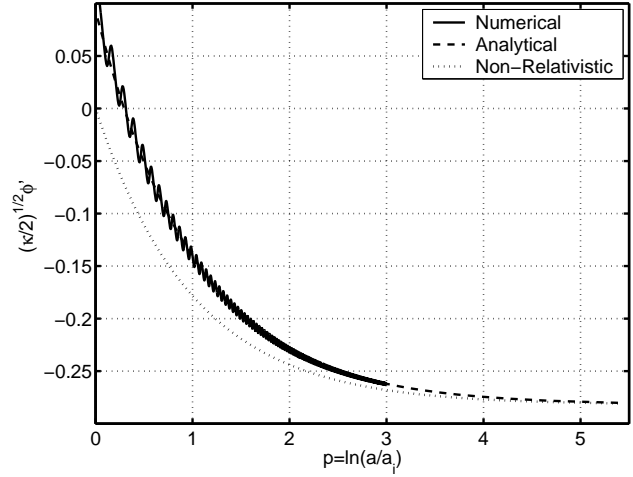}
\end{center}
\caption{Evolution of $\varphi'$ as a function of the p-time variable
in the weak field limit
($a_i=1$, $\rho_{BI}(a_i)/\epsilon_{c}=10^{-3}$, $\epsilon_c=10^{-4}\times
m_{Pl}^4$, $k=1$, $\phi'_i=0.1$, $\delta H/H<10^{-7}$)
}
\label{phip_ym}
\end{figure}

\begin{figure}
\begin{center}
\includegraphics[scale=0.45]{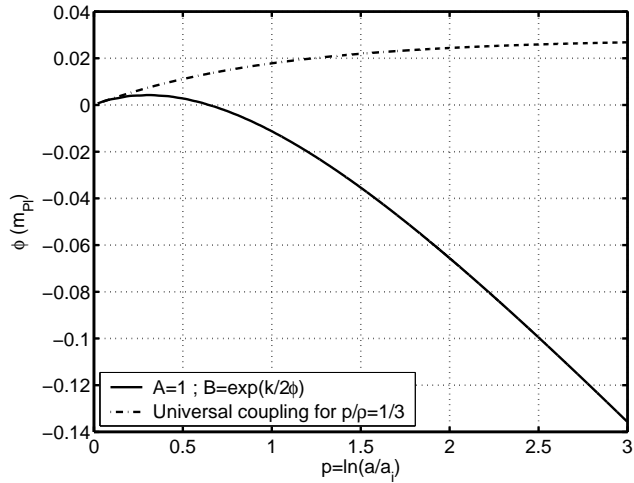}
\end{center}
\caption{Evolution of the dilaton field $\phi$ as a function of the p-time variable
in the weak field limit
(Same parameters as above)
}
\label{phi_ym}
\end{figure}

\begin{figure}
\begin{center}
\includegraphics[scale=0.45]{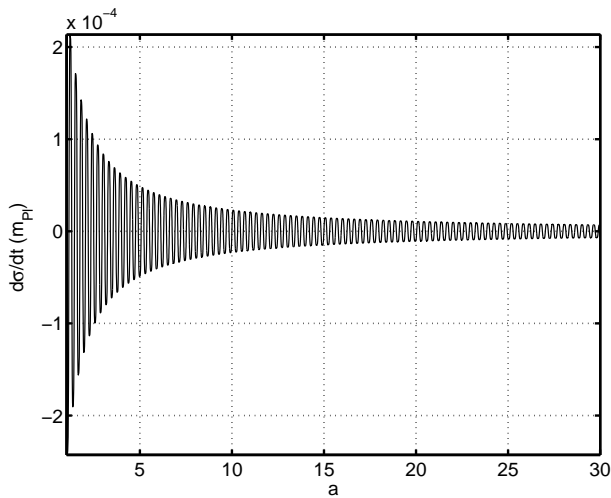}
\end{center}
\caption{Evolution of the gauge field velocity as a function of the
  scale factor in the weak field limit
(Same parameters as above)
}
\label{sigmad_ym}
\end{figure}

The weak field regime of the EBID system is reached when the BI energy
density $A^4(\varphi)\rho_{BI}$ becomes much smaller than the critical energy
$\epsilon_{c}$.
In this case, the lagrangian ruling the gauge sector takes the usual
Yang-Mills form (\ref{lym}), which gives for the spatially homogoneous and
isotropic gauge potentials (\ref{FLRWamu}):
\begin{equation}
\mathcal{L}_{YM}=-\frac{3}{2}\left(\frac{\sigma^4}{a^4}-\frac{\dot{\sigma}^2}{N^2a^2}\right)\cdot
\end{equation}
Taking into account the limit $\epsilon_{c}\rightarrow\infty$ (and
$\mathcal{P}\approx 1$) into the EBID field equations (\ref{hebid}),
(\ref{acc}), (\ref{kg}) and (\ref{bi}) (with $V=0$), we obtain:
\begin{eqnarray}
\left(\frac{\dot{a}}{a}\right)^2&=&\frac{\kappa}{3}\left[\frac{3}{2}A^4(\phi)B^{-4}(\phi)\left(\frac{\dot{\sigma}^2a^2+\sigma^4}{a^4}\right)+\frac{\dot{\phi}^2}{2}\right]\label{h_ym}\\
\frac{\ddot{a}}{a}&=&-\frac{\kappa}{3}\left[\frac{3}{2}A^4(\phi)B^{-4}(\phi)\left(\frac{\dot{\sigma}^2a^2+\sigma^4}{a^4}\right)+\dot{\phi}^2\right]\label{acc_ym}\\
\ddot{\phi}&+&3\frac{\dot{a}}{a}\dot{\phi}-6A^4(\phi)B^{-4}(\phi)\left(\alpha(\phi)-\beta(\phi)\right)\left(\frac{\dot{\sigma}^2a^2-\sigma^4}{a^4}\right)=0\label{kg_ym}\\
\ddot{\sigma}&+&2\frac{\sigma^3}{a^2}+\frac{\dot{a}}{a}\dot{\sigma}+4\dot{\phi}\dot{\sigma}\left(\alpha(\phi)-\beta(\phi)\right)=0\cdot\label{ym}
\end{eqnarray}
These equations constitute the Einstein-Yang-Mills-Dilaton system,
and the special case of $A=1$ and $B(\phi)=\exp(k/2\phi)$ can be found in \cite{bento,bento2}.
In this paper, the authors highlighted
the importance of the energy exchange between the dilaton and the
Yang-Mills field. Indeed, this coupling yields a new force term
in the field equation for the dilaton (\ref{kg_ym}) and the gauge field (\ref{ym}) which
disappear in case of universal coupling ($A=B$ and $\alpha=\beta$). With non-universal coupling
the gravitation is now sensitive to the force term when it is coupled to Yang-Mills
radiation although its equation of state is those of radiation (see (\ref{ts})).
Let us now describe the dynamics of the different fields in the
EBID system for this low-energy regime.\\
\\
First, let us move to the p-time variable $p=\ln (a/a_i)$ and use the acceleration and Friedmann equations
(\ref{acc_ym}), (\ref{h_ym}) to rewrite (\ref{kg_ym})
as
\begin{equation}
\label{wf_phi}
\frac{\varphi''}{3-\varphi^{'2}}+\frac{\varphi'}{3}-2\left(\alpha(\phi)-\beta(\phi)\right)\frac{\dot{\sigma}^2a^2-\sigma^4}{\dot{\sigma}^2a^2+\sigma^4}=0,
\end{equation}
with
$$
\varphi=\sqrt{\frac{\kappa}{2}}\phi\cdot
$$
The energy exchange term in
$\dot{\sigma}^2a^2-\sigma^4 $ inside relations (\ref{kg_ym}) and
(\ref{wf_phi}) is in general oscillating, due to the self-coupling
of the non-abelian gauge field (term in $2\sigma^3/a^2$ in
(\ref{ym})).
A way to handle this easily is to replace it by an effective
source term which would account for the average effect of the gauge
field oscillations. Let us therefore proceed to the following replacement
\begin{equation}
\label{effective}
\frac{\dot{\sigma}^2a^2-\sigma^4}{\dot{\sigma}^2a^2+\sigma^4}\approx \aleph
\end{equation}
with $\aleph$ some constant expressing the effectiveness of the
energy exchange between dilaton and gauge fields in the weak field regime. This constant $\aleph$ can for instance be estimated numerically
by computing the average of the driving term in (\ref{wf_phi}) over one period.
Equation (\ref{wf_phi}) is the same field equation than (\ref{ts}) for the
tensor-scalar theory of the dilaton but now with a non-vanishing force
term due to our averaging of the gauge oscillations.
By averaging the gauge oscillations, we obtain a similar equation
to the strong energy limit (equation (\ref{sf_phi})) seen in
the previous section.
Therefore, we can use the same procedure as
before : if we set $A=1$ and $B(\phi)=\exp(k/2\phi)$, we can propose the following implicit solution for $\varphi'$:
\begin{equation}
\label{sol_ym}
\left(6\mathcal{A}^2-2\right)(p+p_0)=-\sqrt{3}\mathcal{A}\ln\left(\left|\frac{\varphi'+\sqrt{3}}{\varphi'-\sqrt{3}}\right|\right)-
\ln\left(\frac{\varphi^{'2}-3}{\left(3\mathcal{A}+\varphi'\right)^2}\right),
\end{equation}
where $\mathcal{A}=2\aleph k/\sqrt{2\kappa}$. Once again, the p-time
derivative of the dilaton field $\varphi'$ evolves towards the following
attractor
$$
\varphi'(p\rightarrow\infty)=-6\frac{\aleph k}{\sqrt{2\kappa}},
$$
and the maximum value allowed for the dilatonic coupling
constant $k$ for which the dilaton remains relativistic
($\varphi'_\infty\rightarrow -\sqrt{3}$) is
$$
k_{max}=\sqrt{\frac{\kappa}{6\aleph^2}}\cdot
$$
It is important to notice the
opposite sign between the attractors of the strong and weak field regimes which will have important consequences on the
cosmological evolution of the dilaton.
In the non-relativistic
limit, $\varphi^{'2}\le 3$, we find the following solution for the
dilaton
\begin{equation}
\label{nr_ym}
\varphi=-3\mathcal{A}\left(e^{-p}+p-1\right)+\varphi_i
\end{equation}
and we see that the dilaton tends to $-\infty$,
if the dilatonic
coupling constant is positive ($\aleph>0$).
\\
\\
Figure \ref{phip_ym} compares our analytical solution (\ref{sol_ym})
coming from an averaging
approximation  (dashed line) to a numerical solution of the full EBID
system (solid line). In the
YM regime, the velocity of the dilaton $\varphi'$ appears oscillating,
around average values given by the approximation (\ref{sol_ym}) with
$\aleph=1/3$. This value numerically appeared to account for the average behaviour
of the EBID system in the low-energy limit for a wide range of
parameters ($k$, $\epsilon_c$, $\varphi_i$ or $\varphi'_i$).
Therefore, averaging the oscillations of the source term to about a third of their amplitude
seems in very agreement with numerical solutions.
Therefore, the attractor for the
p-time derivative of the dilaton is now
\begin{equation}
\label{att_ym}
\varphi'(p\rightarrow\infty)=-2\frac{k}{\sqrt{2\kappa}},
\end{equation}
while the maximum value allowed for the dilatonic coupling constant is
\begin{equation}
\label{kmax_ym}
k_{max}=\sqrt{\frac{3\kappa}{2}}\cdot
\end{equation}
\\
\\
When $A=1$ and $B(\phi)=\exp(k/2\phi)$, the p-time derivative of the dilaton appears to converge to a constant
negative value which is directly given by the non-relativistic approximation
(\ref{nr_ym})
which is valid when $k$ is small compared with $\sqrt(\kappa)$ (dotted
line). When  $B=1$ and $A(\phi)=\exp(k/2\phi)$, it is obvious from equation (\ref{wf_phi})
that the attractor has exactly the opposite value.
Therefore, the case of non-universal coupling $A\ne B$ is quite different to what happens in a usual tensor-scalar
theory:
in a radiation-dominated
universe,
the p-time velocity of the dilaton
freezes to zero (see the solution
(\ref{phirad})). Indeed, this will make the dilaton field diverging
after an infinite amout of time as can be seen in Figure
\ref{phi_ym}. In this figure, we represented the evolution of the
dilaton in the numerical solution with $A=1$ and $B(\phi)=\exp(k/2\phi)$ (solid line) to the solution
(\ref{phirad})
for the same field in a radiation-dominated universe. More precisely,
the dilaton tends to $-\infty$. As a conclusion, although the gauge field looks like radiation at a large-scale level,
the
non-universal
microscopic coupling between dilaton and gauge sectors finally
dominates at large redshift and freezes the energy contribution of
both sectors in such a way that none of these components completely
dominates.
Figure \ref{sigmad_ym} presents the evolution of the gauge field
velocity $\dot{\sigma}$ with the scale factor. The gauge field appear
to be damped by its coupling to the dilaton and in fact the whole
gauge sector looses energy at a rate fixed by (\ref{cons}).

\begin{figure}
\begin{center}
\includegraphics[scale=0.45]{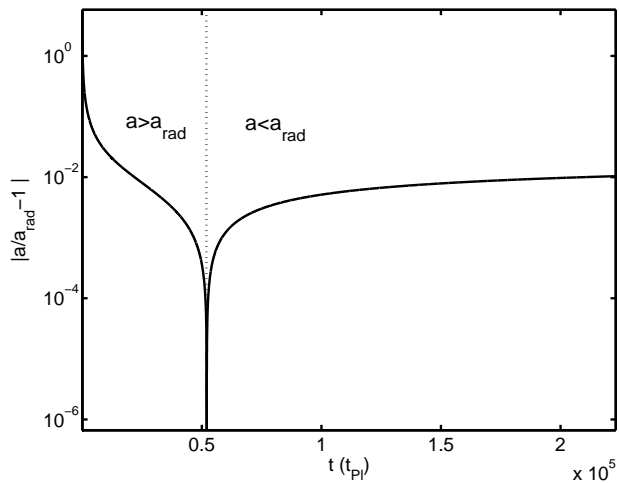}
\end{center}
\caption{Evolution of the departure from a radiation-dominated
  universe of the weak field limit during expansion
(Same parameters as above)
}
\label{a_ym}
\end{figure}

Figure \ref{a_ym} shows the departure of the scale factor from the
radiation solution:
$$
\left|\frac{a}{a_{rad}}-1\right|
$$
for the numerical solution presented in this section. We see that the
departure is important when the scalar field is dominating at early
times ($t<5\times 10^4 t_{Pl}$) then finally converges to a slightly
less strong expansion at late times ($a\approx 0.99\times a_{rad}$),
when the equilibrium has been reached.

Before going further, let us summarise the cosmological evolution of
the fields constituting the EBID system in the YM regime. First, the
dilaton is damped until its velocity is attracted to a negative (resp. positive) value for
the particular coupling  $A=1$ and $B(\phi)=\exp(k/2\phi)$ and $k>0$ (resp. $B=1$ and $A(\phi)=\exp(k/2\phi)$).
However, it should have been damped to rest if it would
have been plunged in a radiation-dominated universe with universal coupling.
Because its velocity has been attracted to a
negative value, the dilaton
field will eventually diverge linearly to $-\infty$ ($k>0$). This is exactly
the opposite situation of the strong field limit that was presented
before where the coupling term drives the dilaton to infinitely high
values.
In a general situation where the gauge field starts with an energy
much higher than the BI critical energy and then cools down to YM
dynamics, one should expect that the dilaton reaches some extremum value ($\varphi'=0$) during
the transition. This will be treated in the next section.
\section{General Cosmological Evolution}
\begin{figure}
\begin{center}
\includegraphics[scale=0.45]{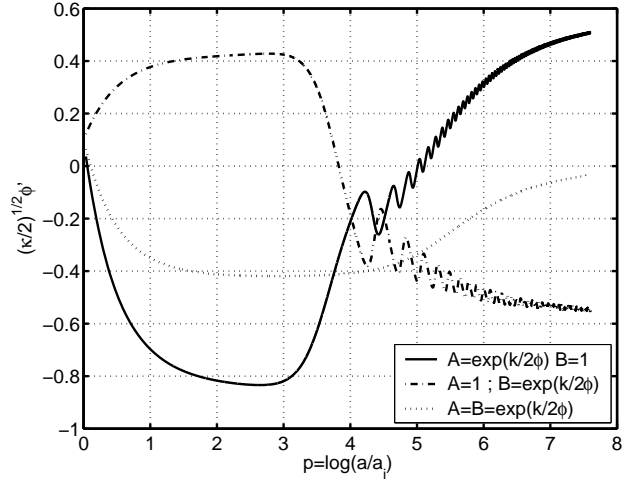}
\end{center}
\caption{Evolution of the dilaton velocity with p-time for three different couplings:
$\beta=0$ and $\alpha=k/2$ (solid line) ; $\alpha=0$ and $\beta=k/2$ (dashed-dot) and universal coupling ($A=B$, dotted line)
($a_i=1$, $\rho_{BI}(a_i)/\epsilon_{c}=10^{4}$, $\epsilon_c=10^{-4}\times
m_{Pl}^4$, $k=2$, $\phi'_i=0.1$, $\delta H/H<10^{-10}$)
}
\label{phip_trans}
\end{figure}

\begin{figure}
\begin{center}
\includegraphics[scale=0.45]{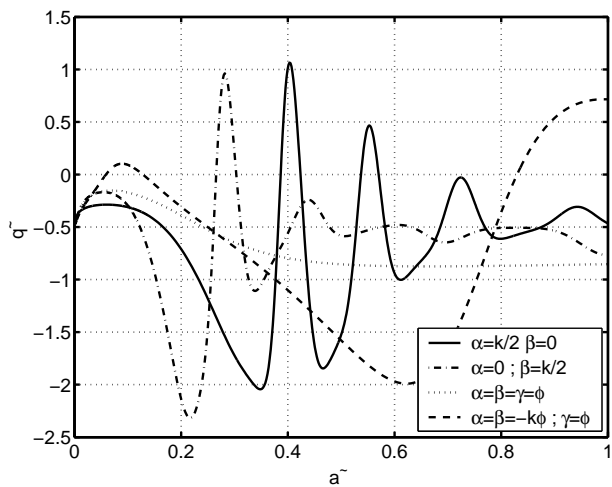}
\end{center}
\caption{Evolution of acceleration parameter in the Jordan frame for four different couplings:
$\beta=0$ and $\alpha=k/2$ (solid line) ; $\alpha=0$ and $\beta=k/2$ (dashed-dot) ; universal coupling ($\alpha=\beta=\gamma=\varphi$, dotted line)
and $\alpha=\beta=k\phi$, $\gamma=\varphi$ (dashed-dashed, $k<0$)
($a_i=1$, $\rho_{BI}(a_i)/\epsilon_{c}=10^{5}$, $\epsilon_c=10^{-4}\times
m_{Pl}^4$, $k=4$, $\delta H/H<10^{-10}$, $\varphi'_i=0.1$, $\rho_m(a_i)/\rho_{BI}(a_i)\approx 100$)
}
\label{qtilde}
\end{figure}

Let us follow in detail some typical cosmological evolutions of the EBID
system for various couplings.
Starting at singularity, the dilaton is in general relativistic
($|\varphi|\rightarrow \infty$ and $\varphi'^2\rightarrow 3$).
The expansion therefore begins at $a=0$ with an
infinite rate and the gauge field dynamics is dominated by the
non-local effects induced by the BI non-linearity ($\rho_{BI}\gg \epsilon_{c}$).
As energy is exchanged between the dilaton and the gauge field in the strong-field regime,
the dilaton velocity is attracted to some value
depending on the coupling functions, as illustrated in figure \ref{phip_trans}.
During this phase, the velocity of the dilaton in p-time is indeed a positive (negative) constant when
$A=1$ and $B(\phi)=\exp(k/2\phi)$ ($B=1$ and $A(\phi)=\exp(k/2\phi)$ or the universal coupling $A=B$).
When the gauge energy density has decreased to the BI critical energy
($\rho_{BI}\approx\epsilon_{c}$),
the dilaton velocity leaves the strong-field attractor to enter the YM
low-energy regime. The epoch of this transition varies according to the coupling functions considered (see Figure \ref{phip_trans}).
It then moves to the low-energy attractor by
accomplishing damped oscillations around the analytical solutions
proposed in the previous section (with $A\ne B$) or is damped to vanishing velocities when there is universal coupling (see section 3).
With non-universal coupling ($A\ne B$), the value of the dilaton field reaches
some extremum ($\varphi'=0$) on its way to the second attractor.
This extremum is unavoidable as we have seen
that the attractors of the dilaton velocity in the strong and weak field regimes are of opposite signs and
its velocity will therefore vanish at some time during the transition
between these two attracting regimes.
\begin{figure}
\begin{center}
\includegraphics[scale=0.45]{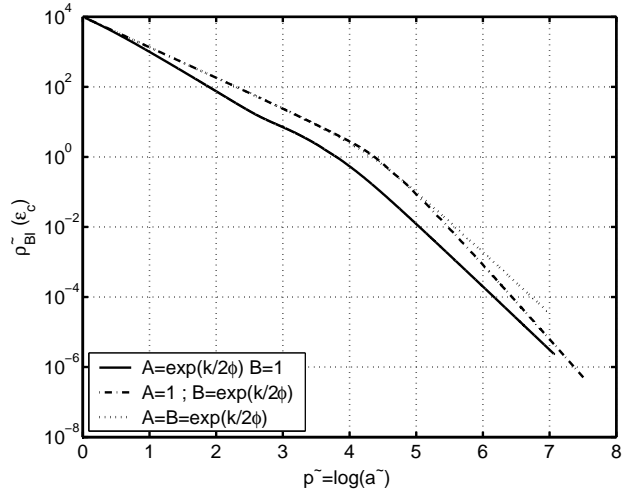}
\end{center}
\caption{Evolution of the physical energy density $\tilde{\rho}$ with $\tilde{p}=\ln(\exp(k/2\phi)a)$ for the curves
in Figure \ref{phip_trans}
(Same parameters as above)
}
\label{rho_tilde}
\end{figure}
It is also interesting to examine the evolution of the gauge field energy density. Figure \ref{rho_tilde}
represents the gauge field energy density $(\mathcal{P}-1)$ related to the curves in Figure \ref{phip_trans}.
With universal coupling $A=B$, we retrieve a cosmological evolution given by equation (\ref{consbis}) and its first integral
(\ref{fi}): $\rho_{BI}\approx a^{-2}$ in the strong field regime and $\rho_{BI}\approx a^{-4}$ at low energies.
The assumption of non-universal coupling now leads to different evolutions of the gauge field energy density, which
in fact are given by the more general energy conservation equation (\ref{cons}). The differences between the evolutions comes from the particular
trajectories illustrated in Figure \ref{phip_trans}.\\
\\
Now that we have reviewed the main features of a general full cosmological
evolution from high to low energy regimes, it is also important to
discuss the evolution of the observed cosmological
parameters like the scale factor, the Hubble expansion rate or the
accelerating parameter in the
Jordan physical frame.
In order to define such a frame, we will introduce an additional matter pressureless fluid which will verify the weak equivalence principle.
The energy density $\rho_m$ of this fluid in the Einstein frame is related to the physical energy density $\tilde{\rho}_m$
through the relation (\ref{jordan2}).
The coupling function $C(\phi)$ to ordinary matter defines now our ``\textit{observable}'' Jordan frame by
\begin{equation}
\tilde{g}_{\mu\nu}=C^2(\phi)g_{\mu\nu}\cdot
\end{equation}
In the Jordan frame obtained by the previous conformal transformation,
the energy density $\tilde{\rho}_m$  of the matter fluid is ruled by the same conservation laws as in general relativity.
This is true because there is no direct interaction between the gauge sector and the additional matter fluid and therefore
they decouple from each other.
The field equation for the gauge potential $\sigma$ does not
need to be modified as we do not assume any direct coupling with the pressureless fluid.
We will now consider the field equations for the EBID system we have written in section 2 for the Einstein metric $g_{\mu\nu}$
with a pressureless fluid $p_m=0$. \\
\\
The observable scale
factor in the Jordan frame will be given by
\begin{equation}
\label{a_jordan}
\tilde{a}=C(\phi)a,
\end{equation}
while the synchronous time in the Jordan frame is denoted by
\begin{equation}
\label{t_jordan}
d\tilde{t}=C(\phi)dt\cdot
\end{equation}
Then, the Hubble expansion rate can be derived directly
\begin{equation}
\tilde{H}=\frac{d\tilde{a}}{\tilde{a}d\tilde{t}}=C^{-1}(\phi)\left(H+\gamma(\phi)\dot{\phi}\right),
\end{equation}
where $H=\dot{a}/a$ is the Hubble parameter in the Einstein frame, a prime denoting a derivative with
respect to $p=\ln(a/a_i)$ (Einstein frame) and $\gamma(\phi)=d\ln C(\phi)/d\phi$.
The
acceleration parameter $\tilde{q}$ in the Jordan frame can be written
\begin{equation}
\label{q_jordan}
\tilde{q}=\frac{\ddot{\tilde{a}}\tilde{a}}{\dot{\tilde{a}}^2}=a\left(\frac{d\gamma(\phi)}{d\phi}\dot{\phi}^2a+\gamma(\phi)\ddot{\phi}a+\gamma(\phi)\dot{\phi}\dot{a}+\ddot{a}\right)\left(\dot{a}+a\gamma(\phi)\dot{\phi}\right)^{-2}
\end{equation}
where a dot over a quantity expressed in the Jordan frame means a derivative with respect to the synchronous time $\tilde{t}$ in that frame.
Although there is no possibility of a cosmic
acceleration ($q>0$)
in the Einstein frame (unless one considers a non-vanishing potential,
see relation (\ref{acc})), this does not rule out a possible
acceleration for the observable scale factor given by
(\ref{a_jordan}).
Indeed, the existence of fluid (constituted by our Born-Infeld non-abelian gauge field)
that violates the weak equivalence principle will result in a possibility of cosmic acceleration.
To show that this is actually the case
even in presence of matter fluid which would make the tensor-scalar theory converging to general relativity if taken alone, we will
use the coupling function $\gamma(\varphi)=\varphi$.
Figure \ref{qtilde} illustrates evolutions of the acceleration
parameter in the Jordan frame
$\tilde{q}(\tilde{p})$, given by
(\ref{q_jordan})
as a function of the Jordan
scale factor $\tilde{a}$. The solutions presented here are expanding universes ($\tilde{H}>0$).
Four different couplings have been considered, including the simple case of $A=B\ne C$ which corresponds
to different couplings of gravitation to the gauge and the matter sector. The pressureless fluid energy density at start has been chosen
to dominate the BI energy density by more than one order of magnitude. As the gauge field starts in the strong field regime, its energy density
will scale approximately with $a^{-2}$, depending on the coupling functions (see also Figure \ref{rho_tilde}). Therefore, the gauge sector
rapidly dominates the energy content of the universe. The dynamics of the dilaton is as described earlier: after having moved to the strong field
regime attractor, the transition to YM dynamics occurs and the dilaton quickly moves to the low-energy attractor. During this transition, acceleration
appears in the Jordan frame defined by the pressureless fluid as indicated in Figure \ref{qtilde}. We see also that any violation of the weak equivalence
principle (here by taking $C\ne A$, $C\ne B$ or $C\ne A=B$) lead to cosmic acceleration even
with tensor-scalar theory that would alone converge to general relativity. An important condition
for cosmic acceleration is to have a repulsive force term in the dilaton equation. Therefore, the EBID system with a non-universal coupling to
gravitation offers the interesting possibility to build a scenario for dark energy or inflation.
For the curves represented here, the acceleration periods are shorter than in a usual LCDM or quintessence model\footnote{To give an idea on how these universes
  accelerate, we remind the reader about the following values of the
  acceleration parameter for various energy content:

\begin{eqnarray}
q(radiation)&=&-1\nonumber\\
q(relativistic\; \phi)&=&-2\nonumber\\
q(\Lambda)&=&1\nonumber\\
q(ghost)&=&2\nonumber
\end{eqnarray}
where $\Lambda$ tends for the cosmological constant (the de Sitter
solution to which inflation
is usually matched as an exponential expansion). }.
and therefore a more complete study
should be done to determine if it is possible to explain distance-redshifts measurements with EBID fields.
It should also be noticed that an EBID dark energy scenario would predict a finite period of acceleration. Indeed,
as the gauge field will recover a YM dynamics at the end of its evolution, its energy density will finally scales
as $a^{-4}$ and will finally be dominated by a pressureless fluid.
Therefore, the
questions whether cosmic acceleration ($\tilde{q}>0$) can occur, with
which intensity and for how long seems
to depend on both
initial energy distribution, the value of the dilaton coupling
constant and the critical BI energy scale $\epsilon_c$.
More work should focus on that point to see if it would be
possible to build a physically relevant quintessence model with the
EBID field equations. However, the perspectives of cosmic acceleration
in the Jordan frame do exist and this
looks particularly interesting for our view of modern cosmology.
\section{Conclusion}
The non-abelian Einstein-Born-Infeld-Dilaton model provides an interesting framework,
motivated by string theory, to study the impact of large scale non abelian gauge fields on tensor-scalar theories of the gravitational
interaction. In this paper, we focused on the cosmological
evolution of an homogeneous and isotropic configuration of these fields in a flat background.
The microscopic coupling between the dilaton and gauge fields induced by non-universal coupling to the metric
leads to an energy exchange between both gravitational and gauge sectors
that will alter the usual dynamics of tensor-scalar theories (for which the coupling is universal).\\
\\
As in non-abelian Born-Infeld cosmology, we considered two different regimes depending
on the gauge energy density compared to the critical energy that parametrizes the Born-Infeld lagrangian.
We have derived both analytical and numerical solutions to describe the cosmological evolution of the whole system.
In the case of non-universal coupling, the gravitational scalar field no more depends on the equation of state
of the gauge field and the dynamics is altered as follows.\\
\\
In the particular case of a Brans-Dicke theory, in which non-perturbative terms for the dilaton are not considered, we have shown that
the energy exchange resulting from the particular couplings
damps the dilaton to a frozen non-vanishing velocity. In the high energy regime of the gauge dynamics,
the attracting value for the velocity is positive when the gauge field couples more to the metric than to the volume form.
The opposite situation happens in the low-energy regime where the gauge field is ruled by Yang-Mills lagrangian.
Therefore, in a general cosmological evolution where the gauge field cools down to low-energies, there is a transition between the two
attractors. Their values are directly proportional to the value of the dilatonic coupling constant. \\
\\
However, it is well-known from
experimental tests of the gravitational theory, especially the determination of the post-newtonian parameter $\bar{\gamma}$, that the
value of the coupling $\omega_0$ is at least of order $500$, roughly $10^{-3}$ for $\alpha_0$
(see \cite{bertotti} for a recent estimation of the post-newtonian parameter $\gamma$).
The influence of the dilaton potential is also important to consider.
Furthermore, the constraints on the weak equivalence
principle obtained by the tests on the universality of free fall exclude a violation of this principle that would exceed a part in $10^{-12}$.
One can therefore argue that the effect of such non-universal couplings
should be neglected.
But, if the violation of the weak equivalence principle only applies to large-scale fields which
do not couple to ordinary matter
and whose distribution is roughly homogeneous,
their energy density on our scales is far beyond experimental reach and the violation could be hard to exhibit.
\\
\\
The interesting possibility introduced by such a violation is a cosmic acceleration in the physical frame associated to ordinary matter.
In this work, we build a first simple model based on our treatment of the EBID field equations that exhibits periods of accelerations in presence
of ordinary matter verifying the weak equivalence principle. The acceleration has been shown to resist to the attracting property of the accompanying
matter and seems to be a general feature of a non-universal coupling to gravitation.
Furthermore, this model respects the weak energy condition $\rho+3p>0$ in the frame of the physical degrees of freedom (the Einstein frame).
The Born-Infeld dynamics of the gauge field plays a crucial role in this kind of dark energy model by ensuring a late arising
of this mechanism (when the gauge field mimics a Nambu-Goto string gas) and even predicts an end to the dark energy domination (when the gauge
field looks like radiation).
An interesting perspective
to this work would be to use this remarkable feature of non-abelian Born-Infeld gauge fields
to build physical models for quintessence and, maybe, inflation.\\
\\
In conclusion, we can say that our study of non-abelian Born-Infeld gauge fields coupled to
to tensor-scalar gravity
opens new and interesting perspectives for the question of the attraction to general relativity as well as other crucial topics of modern cosmology such
as inflation or dark energy.
\section*{Acknowledgements}
The authors warmly thank J. Larena for the many interesting
discussions we had about the underlying physics of this paper.
\section*{Appendix : numerical integration of the EBID system}
Here we give some details of the numerical integration of the full
EBID system we use to illustrate this paper. In order to integrate
the system of equations (\ref{acc}), (\ref{hebid}), (\ref{kg}) and
(\ref{bi}), we choose the following procedure. First, we rewrite
equations (\ref{acc}), (\ref{kg}) and (\ref{bi}) as a system of six
first-order ODE's and keep the hamiltonian constraint (\ref{hebid})
to check the consistency of the numerical computation. Then, let us
redefine the fields in such a way that they will be approximately of
the same order of magnitude (this will avoid stiffness problems in
the integration):
\begin{eqnarray}
a&=&\frac{A}{a_i}\nonumber\\
\phi&=&\Phi m_{Pl}\nonumber\\
\sigma&=&\Sigma m_{Pl},\nonumber\\
\end{eqnarray}
where $A,\; \Phi$ and $\Sigma$ will be the fields to
integrate. It is also useful to set $k=K/m_{Pl}$ and
$\epsilon_c=\epsilon_c' m_{Pl}^4$.
Once the equations have been rewritten under these considerations, we
choose the initial conditions as follows: $a_i$ is set to $1$ and
$\phi_i$ to zero (we therefore start with a ``$bare$'' gauge coupling
constant equal to unity in the Einstein frame). We choose the ratio
$\rho_{BI}(a=a_i)/\epsilon_{c}=r$
so that we can control the type of gauge dynamics (BI, YM or
transition) we start with. Then, we choose the value of
$\phi'(a_i)=\phi'_i$ so that the initial expansion rate will be given
by $H_i^2=\kappa/3\rho_{BI}(a_i)/(1-\kappa\phi^{'2}_i/6)$. This gives also
$\dot{\phi}_i$ as it is equal to $H_i\phi'_i$. Then, without
loss of generality, we can assume $\sigma_i=0$ and determine
$\dot{\sigma}_i$ from the postulated value of $\rho_{BI}(a=a_i)$.
Numerical integration of the system of six ODE's is performed using
the standard method of Shampine-Gordon \cite{gordon}. To monitor the
accuracy of the numerical solution, we compute the absolute violation
of the hamiltonian constraint (\ref{hebid}):
$$
\frac{\delta H}{H}=\frac{\left|H-\dot{A}/A\right|}{H}
$$
all along the integration. The numerical integration makes the violation of the
hamiltonian constraint diverging exponentially with time and we
indicated in the previous figures the final absolute error reached for
each of the numerical solution that were presented.

\label{lastpage}

\end{document}